\documentclass[reprint,amsmath,amssymb,smsymb,aps,superscriptaddress,apl,prd,nofootinbib]{revtex4-2}
\usepackage{geometry}
\renewcommand{\thesubsection}{\alph{subsection}}
\usepackage{titlesec}
\usepackage{bm}
\usepackage{gensymb}
\usepackage{graphicx}
\usepackage{verbatim}
\usepackage{courier}
\usepackage{mathtools}
\usepackage[inline]{enumitem}
\usepackage{calc}
\graphicspath{ { } }
\usepackage{color}
\usepackage{amsmath}
\usepackage{subfigure}
\usepackage{ifthen}
\usepackage{acro}

\definecolor{amber}{rgb}{1.0, 0.49, 0.0}
\definecolor{shamrockgreen}{rgb}{0.0, 0.62, 0.38}

\newboolean{comments}
\setboolean{comments}{false}
\ifthenelse{\boolean{comments}}{
\newcommand{\mdh}[1]{{\textcolor{red}{\sf{[Mark: #1]}}}}
\newcommand{\steve}[1]{{\textcolor{magenta}{\sf{[Steve: #1]}}}}
\newcommand{\charlie}[1]{{\textcolor{blue}{\sf{[Charlie: #1]}}}}
\newcommand{\rhys}[1]{{\textcolor{amber}{\sf{[Rhys: #1]}}}}
\newcommand{\fpg}[1]{{\textcolor{shamrockgreen}{\sf{[Francesco: #1]}}}}
}{
\newcommand{\mdh}[1]{{}}
\newcommand{\steve}[1]{{}}
\newcommand{\charlie}[1]{{}}
\newcommand{\rhys}[1]{{}}
\newcommand{\fpg}[1]{{}}
}

\DeclareAcronym{gw}{
  short = GW ,
  long = gravitational-wave
}
\DeclareAcronym{bh}{
  short = BH,
  long = black-hole
}
\DeclareAcronym{bbh}{
  short = BBH,
  long = binary-black-hole
}
\DeclareAcronym{snr}{
  short = SNR,
  long = signal-to-noise ratio
}
\DeclareAcronym{psd}{
  short = PSD,
  long = power spectral density
}
\DeclareAcronym{pdf}{
  short = PDF,
  long = probability density function
}
\DeclareAcronym{kde}{
  short = KDE,
  long = kernel density estimate
}
\DeclareAcronym{O1}{
  short = O1,
  long = first observing run
}
\DeclareAcronym{O2}{
  short = O2,
  long = second observing run
}
\DeclareAcronym{O3}{
  short = O3,
  long = third observing run
}

\usepackage{booktabs}

\begin{document}

\title{Identifying when Precession can be Measured in Gravitational Waveforms}
\author{Rhys Green}
\affiliation{School of Physics and Astronomy, Cardiff University, Cardiff, CF24 3AA, United Kingdom}
\author{Charlie Hoy}
\affiliation{School of Physics and Astronomy, Cardiff University, Cardiff, CF24 3AA, United Kingdom}
\author{Stephen Fairhurst}
\affiliation{School of Physics and Astronomy, Cardiff University, Cardiff, CF24 3AA, United Kingdom}
\author{Mark Hannam}
\affiliation{School of Physics and Astronomy, Cardiff University, Cardiff, CF24 3AA, United Kingdom}
\affiliation{Dipartimento di Fisica, Universit\`{a} di Roma ``Sapienza'', Piazzale A. Moro 5, I-00185, Roma, Italy}
\author{Francesco Pannarale}
\affiliation{School of Physics and Astronomy, Cardiff University, Cardiff, CF24 3AA, United Kingdom}
\affiliation{Dipartimento di Fisica, Universit\`{a} di Roma ``Sapienza'', Piazzale A. Moro 5, I-00185, Roma, Italy}
\affiliation{INFN Sezione di Roma, Piazzale A. Moro 5, I-00185, Roma, Italy}
\author{Cory Thomas}
\affiliation{School of Physics and Astronomy, Cardiff University, Cardiff, CF24 3AA, United Kingdom}
\date{\today}

\begin{abstract}
In binary-black-hole systems where the black-hole spins are misaligned with the orbital angular momentum, precession effects
leave characteristic modulations in the emitted 
gravitational waveform.
Here, we investigate where in the parameter space we will be able to accurately identify precession, for likely observations over coming 
LIGO-Virgo-KAGRA observing runs. Despite the large number of parameters that characterise a precessing
binary, we perform a large scale systematic study to identify the impact of each source parameter 
on the measurement of precession. We simulate a fiducial binary at moderate mass-ratio, signal-to-noise ratio (SNR), and spins, such that
precession will be clearly identifiable, then successively vary each parameter while holding the remaining parameters fixed.  
As expected, evidence for precession increases with signal-to noise-ratio (SNR), higher in-plane spins, more unequal component masses, 
and higher inclination, but our study provides a \emph{quantitative} illustration of each of these effects, and informs our intuition on 
parameter dependencies that have not yet been studied in detail, for example, the effect of varying the 
relative strength of the two polarisations, the total mass, and the aligned-spin components. 
We also measure the ``precession SNR'' $\rho_p$, which was introduced in Refs.~\citep{Fairhurst:2019_2harm, Fairhurst:2019srr} to 
quantify the signal power associated with precession.  By comparing $\rho_p$ with both Bayes factors and the recovered posterior distributions, 
we find it is a reliable metric for \emph{measurability} that accurately predicts when the detected signal contains evidence for precession. 
\end{abstract}

\maketitle


\section{Introduction}

In September 2015, the first direct detection of \acp{gw} marked the beginning of GW astronomy~\cite{Abbott:2016blz}.
Another 14 detections have been announced by the LIGO Scientific and Virgo collaborations (LVC),
the vast majority of which were due to \ac{bh} mergers~\citep{LIGOScientific:2018mvr, Abbott:2020uma, LIGOScientific:2020stg, Abbott:2020khf, Abbott:2020mjq, Abbott:2020tfl}.
Additional events have also been reported by independent groups~\citep{nitz20202, venumadhav2020new, zackay2019highly, zackay2019detecting}.
These GW observations have already provided significant insights into gravitational physics,
cosmology, astronomy, nuclear physics and fundamental physics (see e.g. Refs.~\cite{SchutzDeterminingHubbleconstant1986,Soares-Santos:2019irc, AbbottGravitationalWavesGammaRays2017,abbott2018gw170817,abbott2018gw170817stochastic,ligo2017gravitational,abbott2019tests,LIGOScientific:2018jsj}).
With an order of magnitude more observations expected over the next 5-10 years, as
the sensitivities of the LIGO~\cite{Aasi:2013wya, TheLIGOScientific:2014jea}, Virgo~\citep{acernese2014advanced} and KAGRA~\citep{aso2013interferometer} detectors improve and additional detectors come online,
GW astronomy from compact-binary mergers has the potential to transform our understanding of gravitational and
fundamental physics~\cite{Sathyaprakash:2019nnu,Bianchi:2018ula,Ford:2019nic}.

Everything we learn from GW \ac{bbh} observations is a consequence of a detailed
parameter estimation analysis that extracts the source parameters of the binary. While some parameters
are extracted with good precision, inspiral dominanted signals show strong correlations between certain parameters which means that they cannot
be measured so accurately, for example correlations between the binary's distance and inclination~\citep{Cutler:1994ys, LIGOScientificCollaborationandVirgoCollaborationPropertiesBinaryBlack2016, Usman:2018imj},
the two masses~\citep{Cutler:1994ys, Poisson:1995ef}, and the mass-ratio and spin components aligned to the binary's orbital angular
momentum~\citep{Poisson:1995ef, Baird:2012cu, farr2016parameter, ng2018gravitational}. As well as studies of the inspiral, work has been done to extract the source properties for high mass signals dominated by the merger ringdown, see e.g.~\cite{Graff:2015bba, Haster:2015cnn, Vitale:2016avz, Yu:2017zgi}.

\begin{figure*}[t]
	\centering
	\includegraphics[scale=0.42]{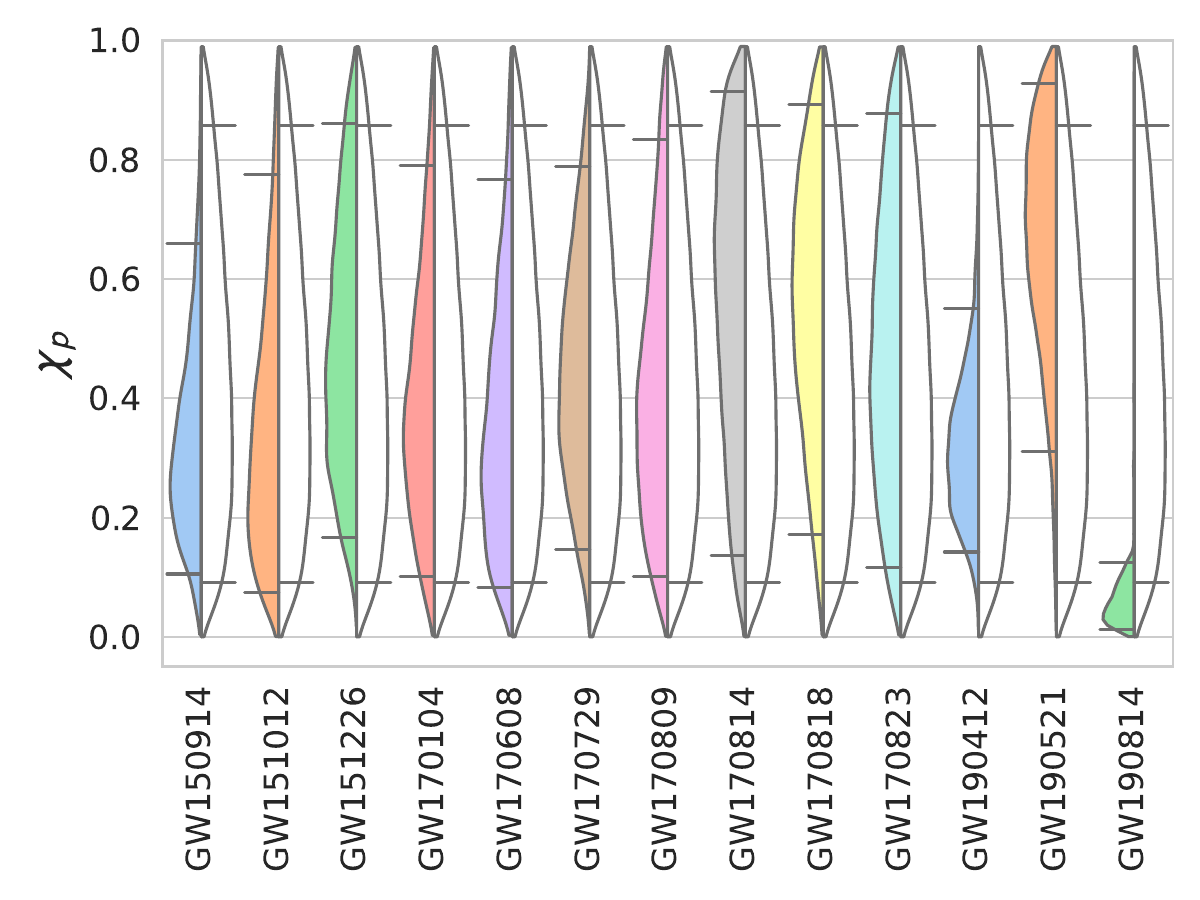}
	\includegraphics[scale=0.42]{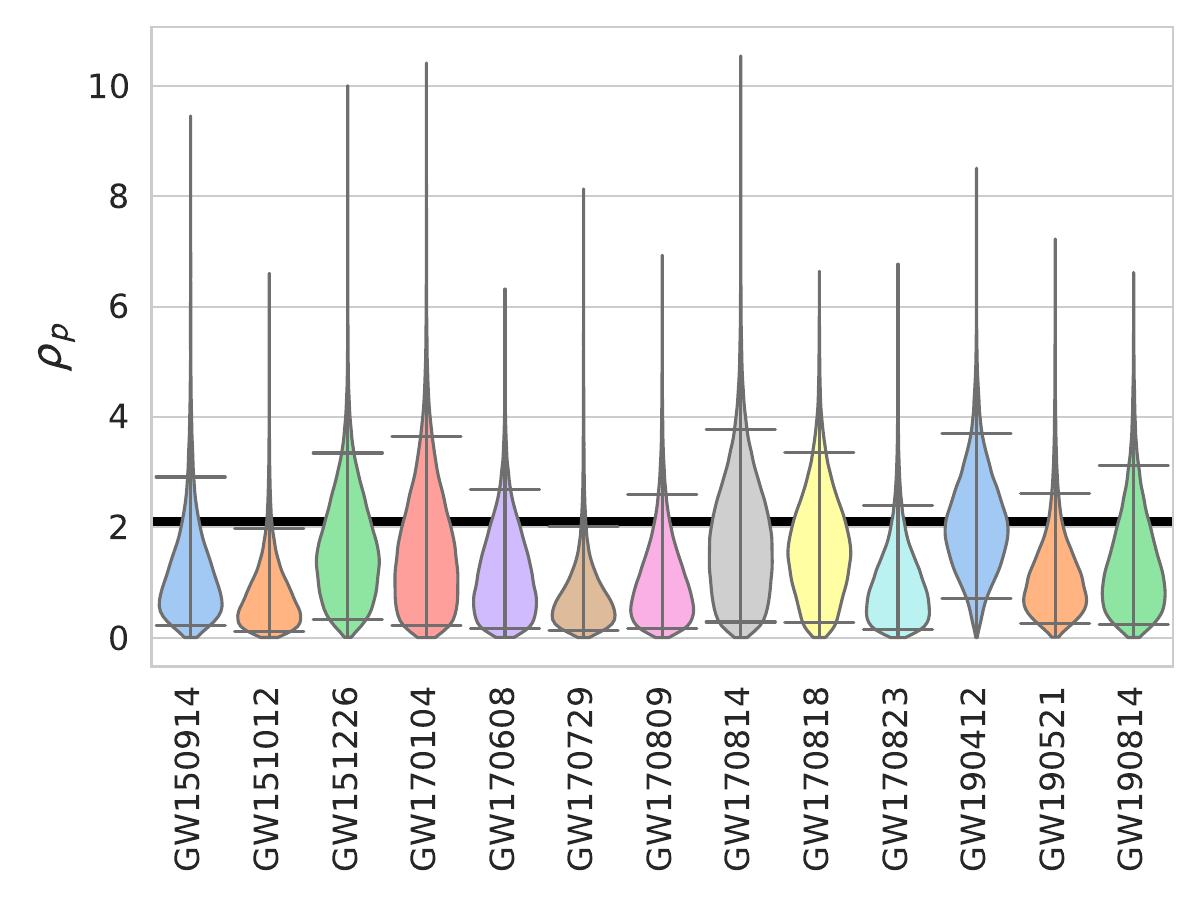}
	\caption{Plot showing the posterior distributions for $\chi_{p}$ and $\rho_{p}$ for all LIGO/Virgo \ac{bbh} observations. The $\chi_{p}$ posterior distribution (left hand side, colored) is compared to its prior (right hand side, white) in the form of a split violin plot. The $\rho_{p}$ posterior distribution is shown as a single violin. Horizontal grey lines show the 90\% symmetric credible interval. The solid black line shows the $\rho_{p} = 2.1$ threshold. Bounded \acp{kde}are used for estimating the probability density. Data obtained from the Gravitational Wave Open Science Center~\citep{abbott2019open}.}
	\label{currentdetections}
\end{figure*}

Spin components misaligned with the binary's orbital angular momentum, leading to a precession of the binary's
orbital plane and hence modulations of the amplitude and phase, have not yet been unambiguously measured
in GW observations~\citep{LIGOScientific:2018mvr}, see Fig.~\ref{currentdetections}. Precession effects and correlations with other parameters are understood in
principle~\cite{Apostolatos:1994mx,Kidder:1995zr} but since theoretical signal models of precessing
binaries that include the merger and ringdown date from only shortly before the first detections~\cite{Hannam:2013oca,Pan:2013rra}, we have less experience of when precession will be measurable, and what the impact will be on other parameter measurements.

The purpose of this paper is to explore when precession will be measurable, and its impact on other parameter measurements, in
the kind of configurations that are representative of expectations from binary populations based on
LIGO-Virgo-KAGRA observations to date~\cite{LIGOScientific:2018mvr}. By utilizing the
precession \ac{snr}  $\rho_{p}$ ~\citep{Fairhurst:2019_2harm, Fairhurst:2019srr} as a
quantifier for the measurability of precession, we also verify that
$\rho_p$ is indeed a good metric for the measurability of precession across the vast majority of the
parameter space, and relate it to the standard means to identify the presence of
precession, the Bayes factor. In doing so, we show that computationally expensive parameter estimation runs can be avoided
by simply calculating the precession SNR.

Previous work has explored the general phenomenology of precession effects: its increased
measurability with large in-plane spins~\citep{vecchio2004lisa, Lang:1900bz, Chatziioannou:2018wqx},
large mass ratios~\citep{vecchio2004lisa, Lang:1900bz},
high inclination~\citep{Apostolatos:1994mx, Brown:2012gs, vitale2014measuring, abbott2017effects, Fairhurst:2019srr, Vitale:2016avz},
and of course high SNR~\citep{vecchio2004lisa, Lang:1900bz, berti2005estimating}.
Beyond these general expectations, the \emph{quantitative} behaviour of parameter measurements in the presence of precession
has not been studied in great detail for typical LIGO-Virgo-KAGRA observations.
The measurability of precession for high mass ratio LIGO-Virgo-KAGRA observations like
GW190814 has been investigated in recent work~\cite{Pratten:2020igi}.

In this paper, we focus on the region of parameter space most likely to yield
binaries with observable precession: binaries of comparable mass, with moderate
in-plane and aligned-spin components ~\citep{Fairhurst:2019srr}.
We perform a series of one-dimensional investigations of the parameter space,
in which we vary one parameter at a time: total mass, mass ratio, spins
(both in-plane as characterized by $\chi_p$, and the aligned spin combination
$\chi_{\mathrm{eff}}$), the binary orientation (both the inclination of the orbit
and also binary polarization), and the sky location and show the impact of
varying each of the binary parameters individually.  These investigations serve
to confirm that much of the known phenomenology is apparent even at relatively low SNR, 
while also demonstrating that the
precession \ac{snr} can be effectively used across
a significant fraction of the parameter space to \textit{predict} the
observable consequences of precession \textit{without} the need for
computationally costly parameter estimation analyses.

This paper is structured as follows: Sec.~\ref{sec:precession} provides an introduction to precession, a brief recap of the two-harmonic
approximation that allows us to define $\rho_p$, and a summary of the importance of precession across the parameter space.
Sec.~\ref{sec:pe_standard} provides an introduction to the parameter estimation techniques used here, and parameter estimation results
and interpretation for our fiducial system.
In Sec.~\ref{sec:pe_results} we perform a series of one-dimensional explorations of the parameter space.
In Sec.~\ref{sec:rho_p_predict} we compare the predicted precession \ac{snr} with observations
and in Sec.~\ref{sec:bayes_factors} we compare precession \ac{snr} with the Bayes factors between precessing and non-precessing runs.
We conclude with a summary and discussion of future directions.


\section{Black hole Spin Induced Precession}
\label{sec:precession}
\begin{figure}[t]
	\centering
	\includegraphics[scale=0.185]{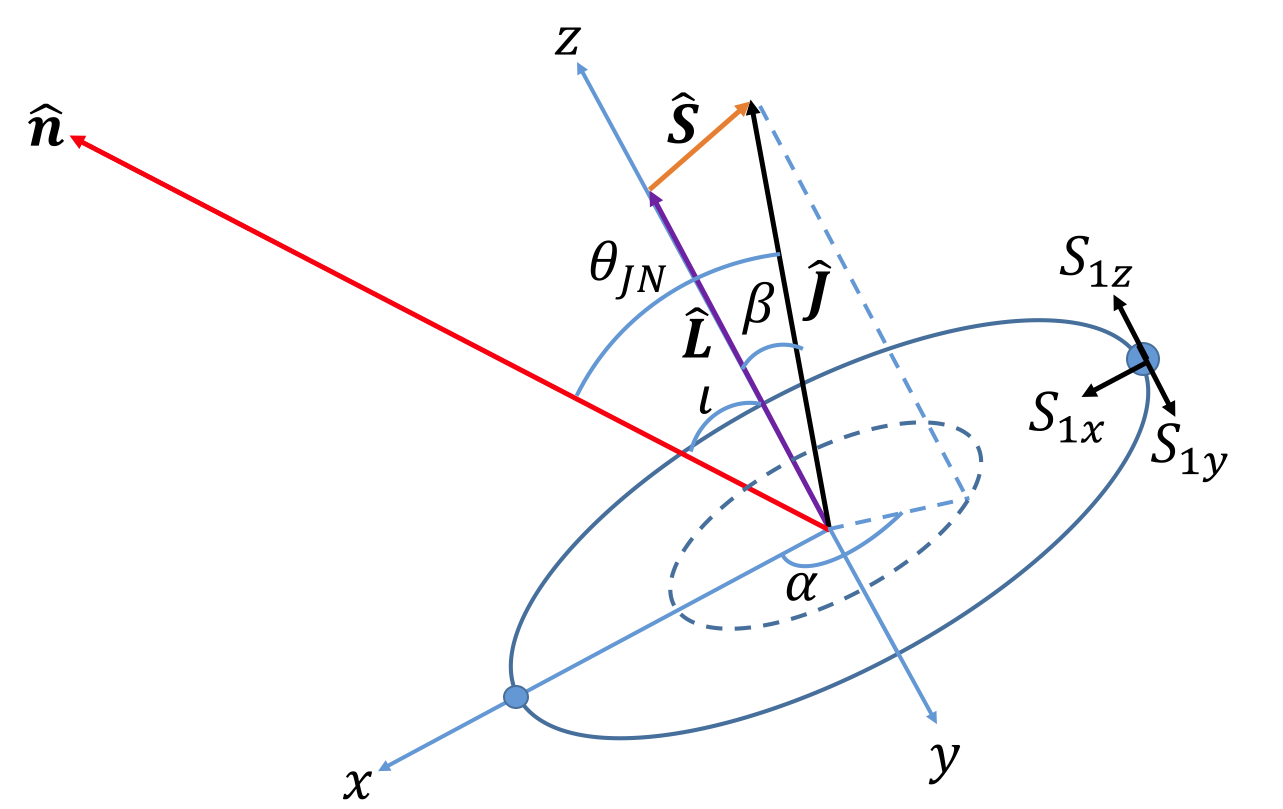}
	\caption{Plot showing how the precession angles used in this study are defined in the co-precessing frame. The normal vector $\mathbf{\hat{n}}$ here indicates the line of sight of the observer, $\mathbf{\hat{L}}$ and $\mathbf{\hat{J}}$ are the orbital angular momentum and total angular momentum vectors respectively, $S_{1x}, S_{1y}$ and $S_{1z}$ are the x, y and z components of the spin on the larger black hole. Note that $\mathbf{J}, \mathbf{L}$ and $\mathbf{\hat{n}}$ are shown to be co-planar only for ease of illustration. It is not true in general.}
	\label{precession_angles}
\end{figure}

A binary consisting of two compact objects will slowly inspiral due to the emission of GWs.
Assuming that the binary is on a quasi-spherical orbit, it may be described by the
individual component masses, $m_{1}$ and $m_{2}$ (where we define $m_1 > m_2$ and we denote
the mass ratio to be $q = m_1/m_2 > 1$), and their spin angular momenta $\mathbf{S}_{1}$
and $\mathbf{S}_{2}$.
\par
For the case where the total spin is misaligned with the total orbital angular momentum,
$\mathbf{S}_{i} \times \mathbf{L} \neq 0$, the system undergoes
spin-induced precession.
In most cases $L \ll J$, and the system undergoes ``simple precession'', where the
orbital angular momentum precesses around the (approximately constant) total angular momentum,
$\mathbf{J} = \mathbf{S_{1}} + \mathbf{S_{2}} + \mathbf{L}$, and the spins precess such that
$\dot{\mathbf{S}} = - \dot{\mathbf{L}}$, where $\mathbf{S} = \mathbf{S_{1}} + \mathbf{S_{2}}$~\cite{Apostolatos:1994mx,Kidder:1995zr}.
\par
The strength of precession is characterised by the tilt angle of the binary's orbit, $\beta$,
defined as the polar angle between $\mathbf{L}$ and $\mathbf{J}$; see
Fig.~\ref{precession_angles}. $\beta$ is determined primarily by the total spin in the plane, and
the binary's mass ratio and separation. At leading order the magnitude of the binary's orbital angular
momentum is given by $L = \mu \sqrt{Mr}$, and we can write,
\begin{equation} \label{eq:opening_angle}
	\cos \beta = \frac{\mu\sqrt{M r} + S_{\parallel}}{\left[\left(\mu\sqrt{M r} + S_{\parallel}\right)^{2} + S_{\perp}^{2} \right]^{1/2}},
\end{equation}
where $M = m_{1} + m_{2}$ is the total mass of the source, $\mu = m_{1}m_{2}/ M $ is the reduced
mass, $r$ is the separation and
$S_{\parallel}$ and $S_{\perp}$ are the total spin parallel and perpendicular to $L$ respectively.
In general, the larger the ``opening angle'', $\beta$, the more prominent the precession effects.

In simple precession cases, $\beta$
slowly increases during inspiral. This temporal variation is mostly dependent on the variation of r as,
 throughout the portion of the binary's inspiral that is visible in the GW detectors, 
 $ \mu, M, S_{\parallel} $ and $ S_{\perp}$ are all approximately constant.
 Therefore, $\beta$ also typically varies very little over the duration that is visible in the GW detectors, 
 and it is possible to make the simplifying assumption that $\beta$ remains constant.
This assumption has been used to good effect in previous work, e.g., Ref.~\cite{Brown:2012gs}, although in this work we make
no assumptions about $\beta$.

It is often convenient to quantify the binary's in-plane spin by the scalar quantity $\chi_{p}$~\cite{Schmidt:2014iyl} (see Refs.~\cite{Gerosa:2020aiw, Thomas:2020uqj} for 
 alternative measures).
$\chi_{p}$ estimates a time-average of the in-plane spin components that drive precession, motivated by the leading-order PN precession
equations~\cite{Apostolatos:1994mx,Kidder:1995zr}, and is defined as,
\begin{equation} \label{chi_p}
	\chi_{\text{p}} = \frac{1}{A_{1}}\text{max}\left(A_{1}S_{1\perp},A_{2}S_{2\perp}\right),
\end{equation}
where $A_{1} = 2+3/(2q)$ and $A_{2} = 2+3q/2$ and $\mathbf{S}_{i\perp}$ is the component of the spin
perpendicular to $\mathbf{L}$ on the $i$th black hole. $\chi_{p}$ as defined above, 
 takes values between $0$ (non-precessing system) and
 $+1$ (maximally precessing system).

\subsection{Two harmonic approximation}
\label{ssec:two_harm}

Non-zero values of $\beta$ and $\chi_p$ indicate that a binary is precessing, and, if all other parameters are kept constant,
they parameterise the amount of precession in a system. However, they are not sufficient to tell us whether precession will
be \emph{observable}. Precession appears in the signal as modulations of the amplitude and phase, but these also depend on the
binary orientation and signal polarisation. Refs~\citep{Fairhurst:2019_2harm, Fairhurst:2019srr} introduce a
method for decomposing a precessing waveform into a series of five {\it{non-precessing harmonics}}, where the
characteristic modulations of a precessing signal are caused by the beating of these {\it{harmonics}}.
The harmonics form a power series in the parameter,
\begin{equation}
b = \tan(\beta/2).
\end{equation}
In most regions of parameter space, the two leading harmonics (a leading-order term independent of $b$, and a first-order
term proportional to $b$)
are sufficient to capture the significant precession features in the waveform, and the other harmonics
can be neglected. As discussed in detail in Sec.~V of Ref.~\cite{Fairhurst:2019_2harm}), 
these other precession harmonics can be ignored for binaries where $\beta \lesssim 45 \degree$.
In general this is true for all binaries, apart from relatively extreme systems, for example, those that have $\theta_{JN}$ close to 
edge-on \emph{and} are either highly precessing or have very large negative spins. 
 
  Thus, for almost all signals we expect to observe, the waveform can faithfully be expressed as,
\begin{equation}\label{eq:h_2harm}
	h(f) \approx \mathcal{A}_{0} h^{0}(f) + \mathcal{A}_{1} h^{1}(f) \, ,
\end{equation}
where $\mathcal{A}_{0}$ and $\mathcal{A}_{1}$ are complex, orientation dependent amplitudes, and $h^{0}(f)$ and $h^{1}(f)$
are the waveforms of the two leading harmonics.  For a detector with a one-sided noise spectral density of $S_n(f)$, the relative amplitude
of the harmonics is given by
\begin{equation}\label{eq:b_bar}
\overline{b} := \frac{ |h^{1} | }{ | h^{0} |} = \sqrt{\frac{\int df \frac{ |h_{1}(f)|^{2}}{S_{n}(f)}}{\int df \frac{ |h_{0}(f)|^{2}}{S_{n}(f)}}} \, .
\end{equation}
which is the average value of $b$  over the frequency range defined by our starting frequency (20Hz) to our Nyquist Frequency (1024Hz), weighted by the signal strength in the detector.
The complex amplitudes $\mathcal{A}_{0,1}$ depend upon the extrinsic parameters of the binary:
the distance $d_L$, angle between $\mathbf{J}$ and the line of sight $\mathbf{\hat{n}}$, polarization angle $\psi$, reference phase $\phi_0$,
and the reference precession phase $\phi_{JL}$\footnote{This is equivalent to $\alpha_{0}$ in Ref.~\citep{Fairhurst:2019_2harm}}.
It is convenient to introduce a reference distance $d_{0}$, which is incorporated into the definitions of the harmonics $h_{0}$ and $h_{1}$.
The amplitudes are then defined as,%
\footnote{These expressions are equivalent to Eq.~(33) in \cite{Fairhurst:2019_2harm}, which are restricted to the
special case where $F_{+} = \cos 2\psi$ and $F_{\times} = -\sin 2\psi$.  More generally, the detector
response also depends upon the sky location in addition to the polarization.}
\begin{align}\label{eq:2harm_amps}
	\mathcal{A}_{0} =& \frac{d_0}{d_L} \left[F_{+} \left(\frac{1+\cos^2 \theta_{\rm JN}}{2} \right) + i F_{\times}\cos \theta_{\rm JN} \right] \times \nonumber \\
	& e^{-i  (2 \phi_{o} + 2 \phi_{JL} )}, \nonumber \\
	\mathcal{A}_{1} =&\frac{d_0}{d_L}  \left[F_{+} \sin 2\theta_{\rm JN}  + 2 i F_{\times} \sin \theta_{\rm JN}  \right] \times \nonumber \\
	& e^{-i (2 \phi_{o} + \phi_{JL} )}.
\end{align}
The relative significance of the two harmonics is encoded by
\begin{eqnarray}\label{eq:zeta}
	\zeta & := & \frac{\overline{b} \mathcal{A}_{1}}{\mathcal{A}_{0}} \\
	& = & \overline{b} e^{i \phi_{JL}}  \, \left(\frac{
	F_{+} \sin2\theta_{\rm JN}  + 2 i F_{\times} \sin \theta_{\rm JN}  }{
	\tfrac{1}{2} F_{+} (1 +\cos^2 \theta_{\rm JN})  + i F_{\times} \cos \theta_{\rm JN}  }\right)  \, , \nonumber
\end{eqnarray}
where the detector response $F_{+, \times}$ is calculated using the polarization angle appropriate for a co-ordinate system defined with the z-axis
along the direction of total angular momentum $\mathbf{J}$.

The observability of precession can then be characterised by the precession \ac{snr} $\rho_{p}$,
defined as the \ac{snr} in the weaker of the two harmonics, and also expressible as a fraction of $\rho = |h|$, the total signal
SNR,

\begin{eqnarray}\label{rho_p}
	\rho_{p} &=& \mathrm{Min}{\left(|\mathcal{A}_{0}h^{0}|, |\mathcal{A}_{1}h^{1}|\right)} \nonumber \\
	&=&  \rho \, \left(\frac{ \mathrm{min}(1, |\zeta| )}{ \sqrt{1 + |\zeta|^{2}}}\right).
\end{eqnarray}

Eq.~\ref{rho_p} assumes that the two harmonics are orthogonal. If the two harmonics are not orthogonal, such that the overlap between the two harmonics is non-zero, it is necessary to consider only the orthogonal part of
the less significant harmonic when calculating the precession SNR. This introduces additional overlap terms between the two harmonics in Eq.~\ref{rho_p}, see Ref.~\cite{Fairhurst:2019_2harm} for details. For simplicity of presentation, in this work all equations are presented under the assumption that the two harmonics are orthogonal, unless otherwise stated. The majority of cases
considered in this paper have a small overlap between the two harmonics (less than 10\%). However, in the analysis we always orthogonalise the harmonics for all calculations. At
higher masses, the two harmonics have significant overlap and we discuss the impact of this when we vary the total mass of the binary below.

The quantity $ \rho_p$ parameterises the \emph{observable} precession,
it is therefore the \emph{measured} quantity in the data.
By considering what we actually measure in the data we are able
to understand many of the correlations and degeneracies in the physical
parameters that are \emph{caused} by the presence of (or lack of)
measurable precession.

In the absence of precession,
$\rho_{p}^{2}$ will be $\chi^{2}$ distributed with two degrees of freedom. Consequently,
 if there is no observable precession in the system, $\rho_{p} \geq 2.1$ in only 10\% of cases. Thus far we have
used $\rho_{p} = 2.1$ as a simple threshold to determine if there is any observable precession in the system.
We revisit this in more detail in Sec.~\ref{ssec:predict_np}.

\subsection{Observability of precession}
\label{ssec:rho_p}

The strength of the modulations in the GW signal depend primarily on the opening angle, $\beta$, and this is reflected in the expansion
parameter $b$ in the two-harmonic approximation; the precession frequency $\dot{\alpha}$ also plays a role.
The strength of the modulations in the \emph{observed} signal also depend on the binary's inclination to the observer, $\theta_{\mathrm{JN}}$, and the
detector polarisation $\psi$, and these are all incorporated into the precession \ac{snr} $\rho_p$, through Eqs.~(\ref{eq:zeta}) and
(\ref{rho_p}). From these we can draw immediate
conclusions about the scenarios in which precession will be most easily measured. These observations are in general not new (see, as always,
the pioneering discussions in Refs.~\cite{Apostolatos:1994mx, Kidder:1995zr}), but we summarise
them here and, where salient, present them in terms of the two harmonic formalism, which highlights the insights and intuition that are
simplified in this formulation. We then compare these expectations with the quantitative results that we find in our full parameter estimation
study.

Our first basic picture of the strength of precession effects comes from Eq.~(\ref{eq:opening_angle}), which gives the dominant
effect on $\beta$ during
the inspiral. If we first consider cases where the spin is entirely in the orbital plane, i.e., $S_{||} = 0$, we see that the opening angle $\beta$ will
be zero if $S_\perp = 0$ (as we would expect), and increases linearly for small $S_\perp$. The opening angle also increases as
$\mu$ decreases, i.e., as the mass ratio is increased. Eq.~(\ref{eq:opening_angle}) is no longer accurate near merger, and for
 equal-mass systems $\beta$ does not become large, but for large mass ratios the opening angle can approach $90^{\circ}$.

If we now consider non-zero $S_{||}$, we see that the level of precession will be reduced for systems with a positive aligned-spin
component, and will be increased for systems with a negative aligned-spin component. The importance of this effect will depend on
the other terms, but we can see that for a high-mass-ratio system where $\mu$ is very small, and close to merger, so $rM$ is also small,
the aligned-spin component will have a strong effect on $\beta$, and therefore the measurability of precession. A negative $S_{||}$ is
necessary to achieve $\beta > 90^{\circ}$, and for large mass-ratio systems near merger (small $\mu$ and $r M$) and large negative
$S_{||}$, $\beta$ can approach $180^{\circ}$, but such systems will be rare.

The measurability of precession also depends on the orientation of the binary with respect to the detector, $\theta_{\mathrm{JN}}$.
As we see in Eq.~(\ref{eq:zeta}), precession effects will be minimal if $\theta_{\mathrm{JN}}\sim 0^{\circ}$ or $180^{\circ}$,
i.e., the observer views the system from the direction of $\mathbf{\hat{J}}$.  We expect precession to be strongest in the observed waveform for orientations close to $\theta_{\mathrm{JN}} \sim 90^{\circ}$.
Additionally, when the detector, or network is primarily sensitive to the $\times$ polarization, precession effects will be more significant.
The amplitude of the $k=1$ harmonic vanishes in the $+$ polarization for both face on $\theta_{\mathrm{JN}}= 0^{\circ}$ and $180^{\circ}$ and edge-on $\theta_{\mathrm{JN}} = 90^{\circ}$ systems,
while the $\times$ polarization is maximal for edge-on systems.  Additionally, the $\times$ polarization for the $k=0$ harmonic vanishes
for edge on systems, while the $+$ polarization is only reduced by a factor of two.  Thus, even when $b$ is small, there can be observable
precession when the system is close to edge on and the network is preferentially sensitive to the $\times$ polarization. For a given
choice of masses and spins, the maximum precession \ac{snr} is $\rho_p = \rho / \sqrt{2}$.


\section{Parameter Estimation Results}
\label{sec:pe_standard}

\subsection{Standard configuration}
\label{sec:standard_inj}
We begin by describing the results of the parameter recovery routine for a specific simulated signal.  The details of the signal are given in Tab.~\ref{tab:standard_run}.  These parameters were chosen so that precession effects would be significant in the observed waveform while still being consistent with the observed population of \ac{bbh}s.  In the following sections, we vary over the parameters of the signal one-by-one to investigate the impact of each parameter on the observability of precession and the accuracy of parameter recovery.  For each parameter, we are able to both increase and decrease the significance of precession.

By taking the inferred properties of the \ac{bbh}s observed in the first, second and third
observing runs~\citep{Abbott:2020niy}, it is predicted that 90\% of detected binaries will have mass ratios
$q < 4$ and $\sim 97\%$ of \acp{bh} in these binaries will have masses less than
$45M_{\odot}$~\citep{Abbott:2020gyp}. 
Our ``standard'' simulated signal was chosen to have total mass $M = 40 M_{\odot}$ and
mass ratio $q = 2$ inclined at an angle of $\theta_{JN} = 60^{\degree}$. This corresponds to component masses of $26.7 M_{\odot}$ and $13.3 M_{\odot}$. This mass ratio and inclination 
was chosen to increase the observability of precession.

Of the 50 events reported by the LIGO/Virgo, 13 exclude the aligned-spin measure $\chi_{\mathrm{eff}} = 0$ at 90\% confidence~\cite{LIGOScientificCollaborationandVirgoCollaborationGW151226ObservationGravitational2016, LIGOScientific:2018mvr, Abbott:2020niy}. The other 37 observations peak at
$\chi_{\mathrm{eff}} = 0$~\citep{LIGOScientific:2018mvr, Abbott:2020niy}. Based on this, studies have shown that
it is likely \acp{bh} in binaries have low spin
magnitudes~\citep{LIGOScientific:2018jsj, Farr:2017uvj, Tiwari:2018qch, Fairhurst:2019srr}. For this reason,
in our standard configuration the \ac{bh} spins were chosen such that there is zero spin aligned with the binary's
orbital angular momentum, $\chi_{\mathrm{eff}} = 0$. We introduce precession by giving the more massive \ac{bh} a
spin of $0.4$ in-plane and leaving the second \ac{bh} with zero spin; two-spin effects are generally far weaker than the
dominant precession effect, which exhibits the same phenomenology as a single-spin 
system~\cite{Buonanno:2004yd,Schmidt:2014iyl}. From Eq.~(\ref{chi_p}) we see that this
gives us a system with $\chi_{p} = 0.4$.
The opening angle for the binary when the signal enters the detector's sensitivity
band is $10^{\circ}$ and the average value of the parameter $b = \tan(\beta/2)$ is $\overline{b}=0.11$, from Eq	.~(\ref{eq:b_bar}).  The
signal is simulated using the \texttt{IMRPhenomPv2} waveform model that incorporates precession effects,
but not higher harmonics ($\ell > 2$) in the signal~\cite{Hannam:2013oca, Bohe:PhenomPv2}.

Our ``standard'' simulated signal was chosen to be more favourable to precession measurements than typical LIGO-Virgo observations. Assuming systems are distributed uniformly in binary orientation, masses drawn from a power law distribution and spins drawn from a low isotropic distribution (see Ref.~\cite{Fairhurst:2019srr} for details), we expect that 4 in every 100 binaries detected by LIGO-Virgo will be inclined at angles greater than $60^{\degree}$ and have $\overline{b} > 0.11$.

The sky location of the binary was chosen to have $\text{RA} = 1.88 \,\text{rad}, \text{DEC}=1.19\, \text{rad}$.
The coalescence time is $t = 1186741861$ GPS (corresponding to the merger time of
GW170814~\citep{abbott2017gw170814}). The polarization angle, defined by the orientation of the orbital
plane when entering the sensitive band at $20\mathrm{Hz}$, is  $\psi = 40\degree$. The two harmonic
approximation is calculated in the J-aligned frame ($\hat{z} = \hat{\mathbf{J}}$). In this frame, the polarization
angle is $\psi_{J} = 120\degree$, which gives antenna factors for H1 of $F_{+}=0.34$ and $F_{\times}=0.53$
and for L1 of $F_{+}=-0.45$ and $F_{\times}=-0.30$, thus both detectors are roughly equally sensitive to
the two GW polarizations.

We injected the signals
into zero noise. The zero-noise analysis results will be similar to those obtained from the average results of
multiple identical injections in different Gaussian noise realisations. The simulated signal is recovered using
the LIGO Livingston and Hanford detectors with sensitivities matching those achieved in the second observing
run (O2) \cite{LIGOScientific:2018mvr}. A low frequency cut-off of
$20$Hz was used for likelihood evaluations, this frequency is also used as the reference frequency when defining 
all frequency dependent parameters such as $\theta_{JN}$.
  Both the LIGO Livingston and Hanford sensitivities improved
prior to the \ac{O3} ~\citep{buikema2020sensitivity} and are expected to improve further prior to the fourth observing
run (O4)~\citep{abbott2018prospects}. The results presented in this work are unlikely to be affected significantly 
by these changes and therefore we expect the main conclusions to be valid for O4 and beyond. 

The \ac{snr} of the signal is
fixed to be 20, corresponding to a moderately loud signal for aLIGO and AdV
observations~\citep{abbott2018prospects}. This sets the distance to $d_L = 223$\,Mpc.
The simulated \ac{snr} in the two
detectors is 16.2 in L1 and  11.7 in H1.  The simulated precession \ac{snr} in each of the detectors is 3.7 and 3.4
respectively, giving a network  precession \ac{snr} of 5.0.  Thus, we expect that precession will be clearly observable in
this signal.

\subsection{Parameter Estimation Techniques}
\label{sec:pe_theory}
We will adopt a parameter estimation methodology that uses matched filtering with phenomenological gravitational waveforms and Markov Chain Monte Carlo (MCMC) techniques to sample the posterior.
\par
We begin by introducing the matched filtering formalisation for parameter estimation. We assume that the time series received from the GW detectors can be decomposed as a sum of the GW signal, $h(t)$, plus noise, $n(t)$, which is assumed stationary and Gaussian with zero mean,
\begin{equation}
	d(t) = h(t) + n(t).
\end{equation}
Under the assumption of Gaussian noise, the probability of observing data $d$ given a signal $h(\bm{\lambda})$ parameterised by $\bm{\lambda} = \{\lambda_{1},\lambda_{2},...,\lambda_{N}\}$, otherwise known as the likelihood, is~\cite{finn1992detection},

\begin{equation} \label{eq:likelihood_calculation}
	p(d|\bm{\lambda}) \propto \exp\left(-\frac{1}{2}\langle d-h(\bm{\lambda}) | d - h(\bm{\lambda}) \rangle \right),
\end{equation}
where $\langle a | b \rangle$ denotes the inner product between two waveforms $a$ and $b$ and is defined as,
\begin{equation}\label{eq:inner_prod}
	\langle a | b \rangle = 4\text{Re}\int_{0}^{\infty}{ \frac{\tilde{a}(f)\tilde{b}^{*}(f)}{S_{n}(f)} df},
\end{equation}
where $S_{n}(f)$ is the one-sided power-spectral density (PSD) and $\tilde{a}$ denotes the Fourier transform of the gravitational waveform $a$.
\par
The posterior \ac{pdf} can then be computed through a simple application of Bayes' theorem,
\begin{equation} \label{bayesequation}
\begin{split}
	p(\bm{\lambda}|d) & = \frac{p(\bm{\lambda})p(d|\bm{\lambda})}{p(d)}, \\
	                & \propto p(\bm{\lambda}) \exp\left(-\frac{1}{2}\langle d-h(\bm{\lambda}) | d - h(\bm{\lambda}) \rangle \right),
\end{split}
\end{equation}
where $p(\bm{\lambda}|d)$ is the posterior distribution for the parameters $\lambda$, $p(\bm{\lambda})$ is the prior probability distribution where $\int{p(\bm{\lambda})d\bm{\lambda}}=1$, and $p(d)$ is the marginalised likelihood where $p(d)=\int{p(\lambda_{i})p(d|\lambda_{i})}d\lambda_{i}$. Posterior distributions for specific parameters can then be found by marginalising over all other parameters,
\begin{equation} \label{posteriorforoneparameter}
	p(\lambda_{i}|d) = \int{p(\boldsymbol{\lambda}|d) d\lambda_{1}...d\lambda_{i-1} d\lambda_{i+1}... d\lambda_{N}}.
\end{equation}

In the idealised situation of zero noise, Eq.~(\ref{eq:likelihood_calculation})
has a maximum at $h(\bm{\lambda})=h(\bm{\lambda}_{0})$. However, as can be seen in Eq.~(\ref{bayesequation}) the posterior also includes priors, this means that, as well as effects due to noise, certain priors may cause the
maxima to be deflected away from $h(\bm{\lambda})=h(\bm{\lambda}_{0})$. This would then lead to Eq.~(\ref{posteriorforoneparameter}) recovering a
biased posterior. In this work, we consider the effect of three closely related priors, 

\begin{itemize}
	\item{\emph{Global}: the prior used during the parameter estimation analysis. This reflects our prior belief  before observing any data,}
	\item{\emph{Conditioned}: the global prior conditioned upon the posterior distributions of other parameters from the same analysis. For example since $\chi_{\mathrm{eff}}$ and $\chi_{\mathrm{p}}$ are correlated, any informative measurement of $\chi_{\mathrm{eff}}$ modifies our prior beliefs about $\chi_{\mathrm{p}}$. This prior has been used in previous LVC publications, see e.g.~\cite{LIGOScientific:2018mvr},}
	 	\item{\emph{Informed}: the global prior conditioned upon the posterior distributions from a different analysis. Here, we use this to inform our expectations of the degree of precession given the results from a non-precessing analysis. See Section~\ref{sec:rho_p_predict} for details.}
\end{itemize}

\subsection{Parameter recovery}
\label{sec:standard_pe}
\begin{table*}[t]
    \begin{ruledtabular}
        \begin{tabular}{ l | c | c c | c c |}
            & Simulated &  \multicolumn{2}{c |}{Median} & \multicolumn{2}{c |}{maxL} \\
            \hline
           & Precessing & Non-Precessing & Precessing & Non-Precessing & Precessing \\
            Total mass $M / M_{\odot}$ & $40.0$ & $40^{+3}_{-2}$ & $40^{+4}_{-2}$ & $40.161$ & $40.507$ \\
            Chirp mass $\mathcal{M} / M_{\odot}$ & 16.22 & $16.5^{+0.3}_{-0.2}$ & $16.3^{+0.3}_{-0.3}$ & $16.459$ & $16.113$ \\
            Mass ratio $q$ & $2.0$ & $1.8^{+0.8}_{-0.7}$ & $1.9^{+0.9}_{-0.7}$ & $1.895$ & $2.191$ \\
            Inclination angle $\theta_{JN} / {}^{\circ} $ & $60.0$ & $110^{+50}_{-100}$ & $120^{+40}_{-90}$ & $30.0$ & $40.0$ \\
            Precession phase $\phi_{JL} / {}^{\circ}$ & $45.0$ & -- & $200^{+100}_{-200}$ & -- & $80.0$  \\
            Effective aligned spin, $\chi_{\mathrm{eff}}$ & $0.0$ & $0.044^{+0.099}_{-0.084}$ & $-0.005^{+0.098}_{-0.092}$ & $0.06$ & $-0.011$ \\
            Effective precessing spin, $\chi_p$ & $0.4$ &  -- & $0.5^{+0.4}_{-0.3}$ & -- & 0.554 \\
            Right ascension $\mathrm{RA} / \mathrm{rad}$ & $1.88$ & $3^{+3}_{-3}$ & $3^{+3}_{-3}$ & $1.418$ & $1.325$ \\
            Declination $\mathrm{DEC} / \mathrm{rad}$ & $1.19$ & $0.2^{+1.0}_{-1.2}$ & $0.2^{+1.0}_{-1.2}$ & $1.229$ & $1.221$ \\
            Luminosity distance $d_{L}/\mathrm{Mpc}$ & $223$ & $500^{+200}_{-200}$ & $400^{+200}_{-200}$ & $451.834$ & $372.706$ \\
            Network \ac{snr} $\rho$ & $20.0$ & $19.3^{+0.1}_{-0.2}$ & $19.7^{+0.2}_{-0.2}$ & $19.52$ & $19.936$ \\
            Precessing \ac{snr} $\rho_{p}$ & 5.05 & -- & $4^{+2}_{-2}$ & -- & $4.649$ \\
            \end{tabular}
        \end{ruledtabular}
    \caption{Table showing the simulated and inferred parameters for the ``standard'' injection when
    recovered by a non-precessing (\texttt{IMRPhenomD}) and a precessing (\texttt{IMRPhenomPv2}) waveform model.
    We report the median values along with the 90\% symmetric credible intervals and the maximum
    likelihood (maxL) value.
    }
    \label{tab:standard_run}
\end{table*}

We performed parameter estimation on the signal using the \texttt{LALInference}~\cite{Veitch:2014wba}
and \texttt{LALSimulation} libraries within \texttt{LALSuite}~\citep{LALSuite}. Parameter recovery was performed with the \texttt{IMRPhenomPv2} model~\citep{Bohe:PhenomPv2, Hannam:2013oca},
which matches the simulated signal to remove any systematic error caused by waveform uncertainty,
and the corresponding \texttt{IMRPhenomD} aligned-spin waveform model~\citep{Khan:2015jqa, Husa:2015iqa},
which does not include any precession effects. Additionally, all analyses used exactly the same priors as those used in the LIGO-Virgo discovery papers, for details, see Appendix B.1 of~\citep{LIGOScientific:2018mvr}. All post-processing was handled by the \texttt{PESummary} python
package\citep{Hoy:2020vys}.

Tab.~\ref{tab:standard_run} summarises the key results for the standard configuration.
All uncertainties are the 90\% symmetric credible intervals.

\begin{figure}[t]
	\centering
	\includegraphics[width=0.49\textwidth]{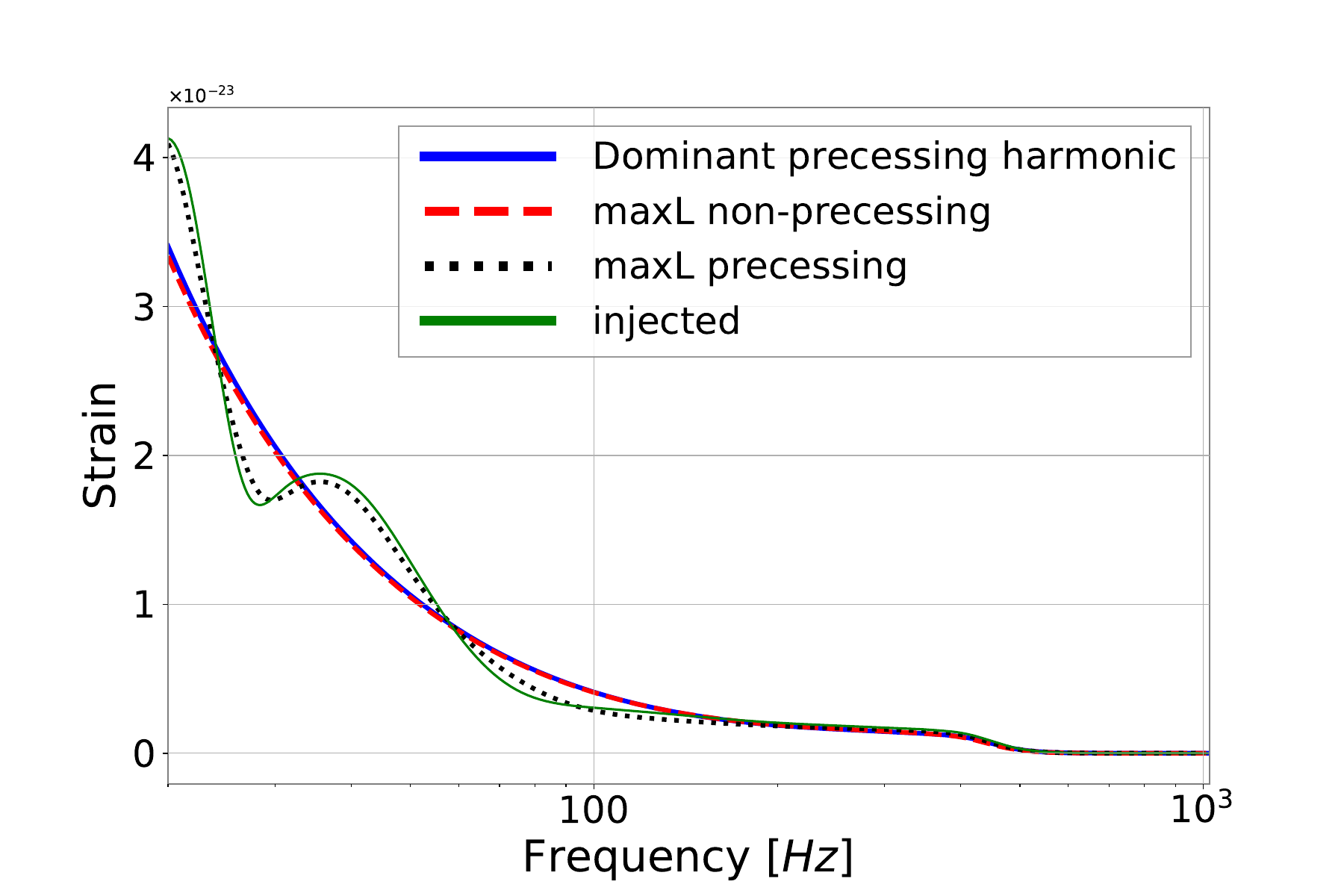}
	\caption{Comparison of the simulated precessing (green), non-precessing maximum likelihood (red), precessing maximum likelihood (black) and dominant precessing harmonic (blue) waveforms as a function of frequency. Waveforms are projected onto the LIGO Hanford detector.
  }
	\label{fig:phenomd_vs_dominant}
\end{figure}

We begin by comparing the overall differences between parameter recovery with the precessing, \texttt{IMRPhenomPv2}, and non-precessing, \texttt{IMRPhenomD}, runs.  From the table, we see that the maximum likelihood \ac{snr} for the non-precessing model is, as expected, lower than for the precessing waveforms.  This can be easily understood from the two-harmonic approximation.  Since the precessing waveform is well approximated by the sum of two non-precessing harmonics, we would expect the non-precessing recovery to accurately recover the more significant of these two.  If that were the case, the we would expect that,
\begin{equation}
	\rho_{D}^{2} \approx \rho^{2} - \rho_{p}^{2},
\end{equation}
and this is indeed the case, as $\rho_{D} = 19.52$, $\rho = 19.94$
and the recovered power in the second harmonic is $\rho_p = 4.6$.
Furthermore, we see that the recovered waveforms confirm this expectation: the recovered waveform when we
include precession matches well with the simulated signal, while the non-precessing run recovers a waveform that
 matches the dominant harmonic, as show in Fig.~\ref{fig:phenomd_vs_dominant}.

\begin{figure*}[t]
	\centering
	\includegraphics[width=0.49\textwidth]{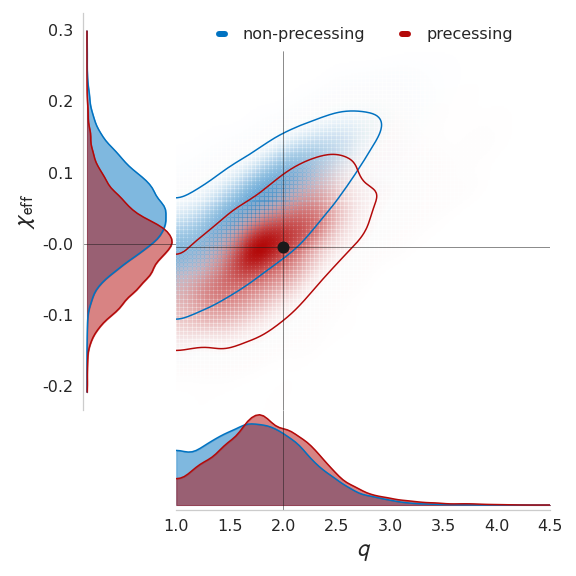}
	\includegraphics[width=0.49\textwidth]{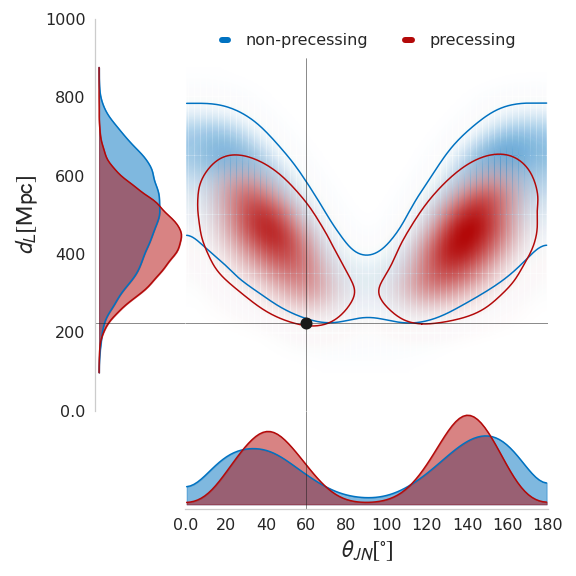}
	\caption{2d contour comparing $q$--$\chi_{\mathrm{eff}}$ (left) and distance--inclination (right) degeneracies when precession effects are included.
  Contours show the 90\% confidence interval. Bounded two-dimensional \acp{kde} are used for estimating the joint probability density.  The black circle indicates the simulated values.}
	\label{fig:p_vs_d_degeneracy_contour}
\end{figure*}

We first consider the accuracy with which the masses and (aligned) spins are recovered.  As expected, the chirp mass
of the system is well recovered, in that it matches the simulated value with only a 2\% uncertainty, which remains constant for both precessing and non-precessing runs.  As is well known, there is a degeneracy between mass-ratio and spin,
particularly during the inspiral part of the waveform \cite{Poisson:1995ef, Baird:2012cu, farr2016parameter, ng2018gravitational}, which leads to significant uncertainty in both
parameters.  In Fig.~\ref{fig:p_vs_d_degeneracy_contour} we show the recovery of the mass ratio and spin, for both precessing and non-precessing runs.  When the model used to recover includes precession effects, the peak of the posteriors is located close to the simulated value ($\chi_\mathrm{eff} = 0$ and $q = 2.0$) and, while the degeneracy leads to significant uncertainty in both parameters, the mass-ratio distribution is clearly peaked away from $q=1$.  Interestingly, when we recover with a non-precessing waveform model, the inferred \textit{aligned} spin component is systematically offset, with a peak at $\chi_{\mathrm{eff}} \approx 0.05$.  This can be understood by recalling that precession induces a secular drift in the phase evolution of the binary, and this can be mimicked by a change in the value of the aligned spin \cite{Apostolatos:1994mx, Fairhurst:2019_2harm}.
This discrepancy has not been seen in LIGO/Virgo observations~\citep{LIGOScientific:2018mvr} as we have
not observed any systems with significant $\rho_{p}$ (see Fig.~\ref{currentdetections}). We investigate this
further in Sec.~\ref{ssec:q_chi}, where we study the effect of varying the mass ratio.

For non-precessing binaries, it is generally not possible to accurately recover the distance and orientation of the source, due to a well known degeneracy (see e.g., Ref.~\cite{Usman:2018imj} for details), although the observation of higher signal harmonics can break this degeneracy through an independent measurement of the source inclination~\citep{Cutler:1994ys, Usman:2018imj, kalaghatgi2019parameter}.  Similarly, the observation of precession can break this degeneracy~\citep{vitale2018measuring}. Precession causes an oscillation of the orbital plane leading to a time-dependence of the orientation of the orbital plane relative to the line of sight.  Equivalently, in the two-harmonic picture, precession leads to the observation of a second harmonic and, consequently, additional constraints on the binary orientation as the amplitudes of the harmonics depend upon the viewing angle. In
Fig.~\ref{fig:p_vs_d_degeneracy_contour}, we show the inferred two-dimensional distance and inclination posteriors for the precessing and non-precessing runs.
As expected, the precessing run constrains the source to be away from face-on,
while the non-precessing run simply returns the prior.  However, even with observable precession, the simulated distance and orientation are not accurately recovered --- a significant fraction of the posterior support is for a system at a greater distance and oriented closer to face-on. We will see how these measurements improve with stronger precession
in Sec.~\ref{ssec:chi_p}.

The sky location of the source is not well recovered.  The analysis was performed with only the two LIGO detectors, and therefore we expect to recover the source restricted to a ring on the sky, which corresponds to a fixed time delay between the detectors \cite{Fairhurst:2009tc, Singer:2015ema}.  The location along the ring cannot be well constrained and, as expected the inferred location is preferentially associated with sky positions where the detector network is more sensitive.  Thus, while the simulated sky location is within the 90\% region, it is not at or close to the peak.  This impacts the recovery of the distance, with the signal being recovered at larger distances, although the simulated distance remains within the 90\% range.  In Section \ref{ssec:sky_loc}, we show results from a set of runs with varying sky location, and verify that at sky locations where the network is more sensitive, the distance posterior is more consistent with the simulated value.

\begin{figure}[t]
	\centering
	\includegraphics[width=0.49\textwidth]{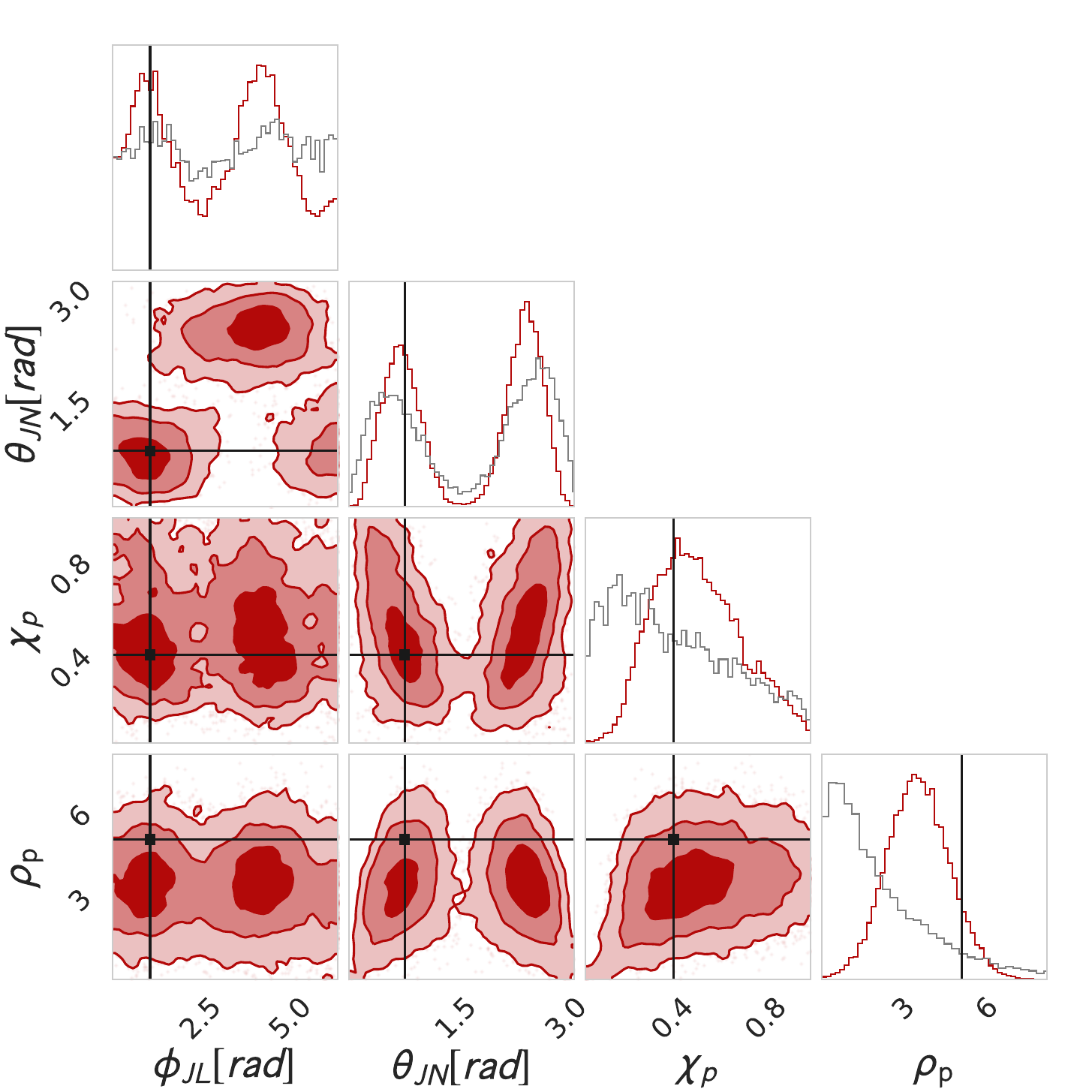}
	\caption{A corner plot showing the recovered values of binary orientation $\theta_{JN}$, precessing spin $\chi_{p}$, precession phase $\phi_{JL}$ and precession \ac{snr} $\rho_{p}$.  Shading shows the $1\sigma$, $3\sigma$ and $5\sigma$ confidence intervals. Black dots show the simulated values. The grey histograms show the \emph{informed} prior, see Sec.~\ref{sec:rho_p_predict}. There is a clear correlation between the binary orientation and inferred precession spin, with signals which are close to face on ($\cos \theta \approx \pm 1$) having larger values of precessing spin, while those which are more inclined having less precessing spin.  The precession \ac{snr} only weakly correlated with $\chi_p$.}
	\label{fig:rho_p_chi_p}
\end{figure}

Lastly, we turn to measurement of precession.  In Fig.~\ref{fig:rho_p_chi_p} we show the recovered distributions for
binary orientation, $\theta_{JN}$, precessing spin $\chi_{p}$, initial precession phase, $\phi_{JL}$, and precession
SNR, $\rho_{p}$.
There is a clear correlation between the inferred orientation and $\chi_p$, with binaries that are more
inclined having lower values of $\chi_p$.  Neither of these quantities are directly observable, it
is only the amount of observable precession in the system, encoded by $\rho_p$, that can be measured.  Thus the orientation
and spin must combine to give the right amount of power in precession, and we see that this is the
case --- there is little correlation between the recovered values of $\rho_p$ and the precessing spin $\chi_p$.
The inferred value of the precessing spin $\chi_{p}$ and precession \ac{snr} $\rho_p$ are both consistent with the
simulated values.  Specifically, the signal has $\chi_p = 0.4$ and this is consistent with the recovered value, although the
posterior distribution is broad, with support over essentially the entire range from 0 to 1. The precession \ac{snr} peaks
well away from zero, giving clear indication of precession in the system. However, the peak of the distribution
occurs at 3.5, while the simulated value is 5.0.  We have deliberately chosen an event with
significant observable precession. Only a small fraction of the parameter-space volume leads to such significant
precession as shown by the \emph{informed} prior on Fig.~\ref{fig:rho_p_chi_p}. This is calculated by estimating the allowed values of $\rho_{p}$ conditioned on the measurements from a non-precessing
analysis. See Sec.~\ref{sec:rho_p_predict} for further details.

The precession phase, $\phi_{JL}$, while not measured with great accuracy, does show two peaks, which are
consistent with the simulated value of $45\degree$ ($0.8$ rad).  The precession phase can be inferred from the
relative phase of the two precessing harmonics using Eq.~(\ref{eq:zeta}), provided the binary
orientation is well measured.  There is a clear dependence with the binary orientation:
if $\theta_{JN} < 90\degree$ then the peak is in $\phi_{JL}$ at the simulated value and if it is greater then $\phi_{JL}$ is offset by
$180\degree$, to compensate for the change in sign of the $\cos \theta_{JN}$ terms in Eq.~(\ref{eq:zeta}).


\section{Impact of Varying Parameters}
\label{sec:pe_results}
We now look at the effect of varying individual parameters one at a time on the recovered posteriors,
in particular focusing on the measurement of precession as described by the posterior distributions of
$\rho_p $ and $\chi_p$. All subsequent one-dimensional investigations of the parameter space maintain a 
constant \ac{snr} (except for Sec.\ref{ssec:snr} where the effect of the \ac{snr} is investigated). This is achieved by 
varying the distance to the source.

Primary results presented in this section will be displayed in the form of violin plots.
We show the $\chi_{p}$ posterior distribution (left hand side, colored) compared to the global prior (right hand 
side, white) unless otherwise stated. We show the $\rho_{p}$ posterior distribution as a single violin. Horizontal 
grey lines show the 90\% symmetric credible interval. Horizontal red lines show the simulated value. A solid 
black line corresponds to the $\rho_{p} = 2.1$ threshold. Bounded \acp{kde}are used for 
estimating the probability density. We use the same 2d contour plots and multi-dimensional corner plots as 
described in Sec.~\ref{sec:standard_pe}. Plots were generated with the PESummary~\cite{Hoy:2020vys} 
python package.


\subsection{SNR}
\label{ssec:snr}
We start with the fiducial run configuration described above and vary the
\ac{snr} of the simulated signal.

In the strong-signal limit, where the likelihood surface can be well approximated by 
a multivariate gaussian, it is well known that the accuracy with which parameters 
can be measured is generally
inversely proportional to the SNR~\citep{Cutler:1994ys, Poisson:1995ef}.  However,
this is not always the case due to, for example, degeneracies between parameters (see
Ref.~\cite{Vallisneri:2007ev} for a discussion of the limits of this approximation).

\begin{figure*}[t]
	\centering
	\includegraphics[scale=0.4]{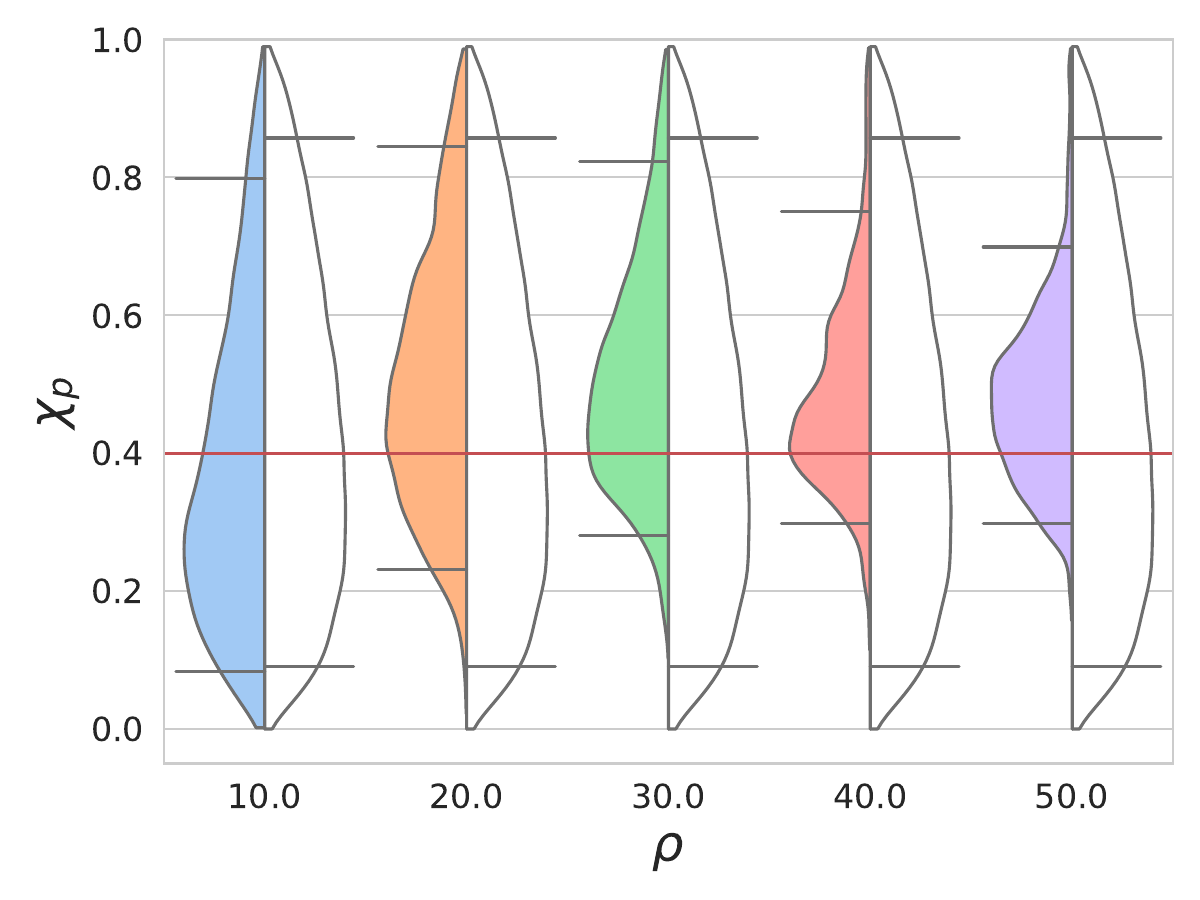}
	\includegraphics[scale=0.4]{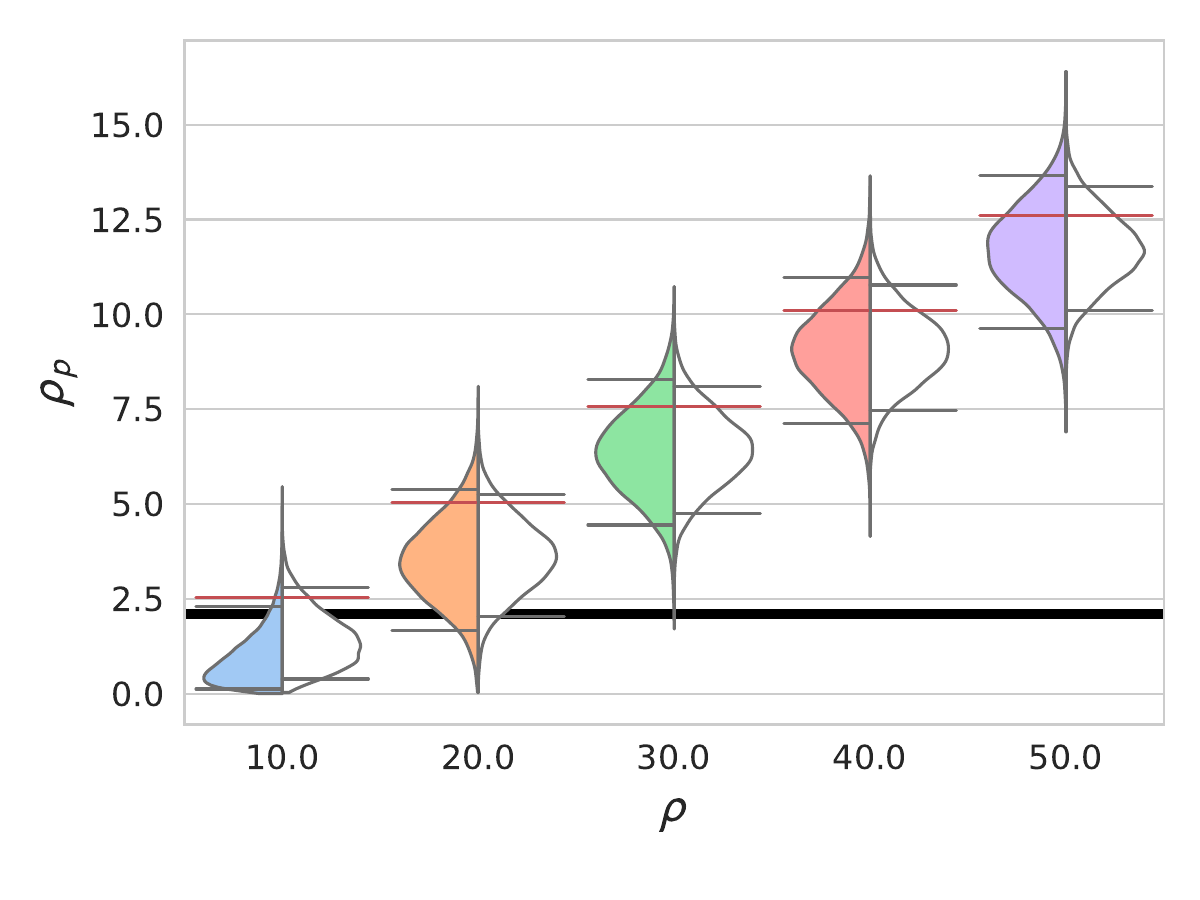}
	\caption{ Violin plots showing the recovered posterior distributions distributions for $\chi_p$ compared to its prior (left) and $\rho_p$ compared to a non-central $\chi$ distribution with 2 degrees of freedom and non-centrality equal to the median of the $\rho_p$ distribution (right). Distributions are plotted for varying SNR. Parameters other than the \ac{snr} of the signal match the ``standard injection'' (see Table~\ref{tab:standard_run}).}
	\label{chi_p_vs_SNR}
\end{figure*}

Fig.~\ref{chi_p_vs_SNR} shows that as the \ac{snr} of the simulated signal increases, the accuracy and
precision of the inferred $\chi_{p}$ posterior distribution improves. As expected the width of the 90\%
credible interval decreases approximately linearly with increasing SNR. The improvement in the $\chi_{p}$
posterior distribution can be mapped to a linear increase in $\rho_{p}$.

When the simulated signal has low \ac{snr} ($\rho = 10$), the recovered $\chi_{p}$ posterior distribution
resembles the prior, implying that there is no information about precession in the data. For this case,
$\rho_{p}$ matches the expected distribution in the absence of any measurable precession --- a $\chi$
distribution with 2 degrees of freedom. As the \ac{snr} increases ($\rho = 20$-$30$), the 5th percentile of the
the $\rho_{p}$ distribution is comparable or greater than the $\rho_{p} = 2.1$ threshold. This maps
to the $\chi_{p}$ posterior distribution removing all support for near-zero $\chi_{p}$ ($\chi_{p}\lesssim 0.1$).
For larger SNRs ($\rho > 40$), the entire $\rho_{p}$ distribution is greater than the 2.1 threshold. This implies
significant power from precession. For these cases, we remove support for maximal precession
$\chi_{p} \sim 1$.

As expected we find good agreement between $\rho_{p}$ and a non-central $\chi$ distribution with 2 degrees
of freedom and non-centrality equal to the inferred power in the second harmonic (median of the $\rho_{p}$
distribution).


\subsection{In-plane spin components}
\label{ssec:chi_p}
\begin{figure*}[t]
	\centering
	\includegraphics[scale=0.4]{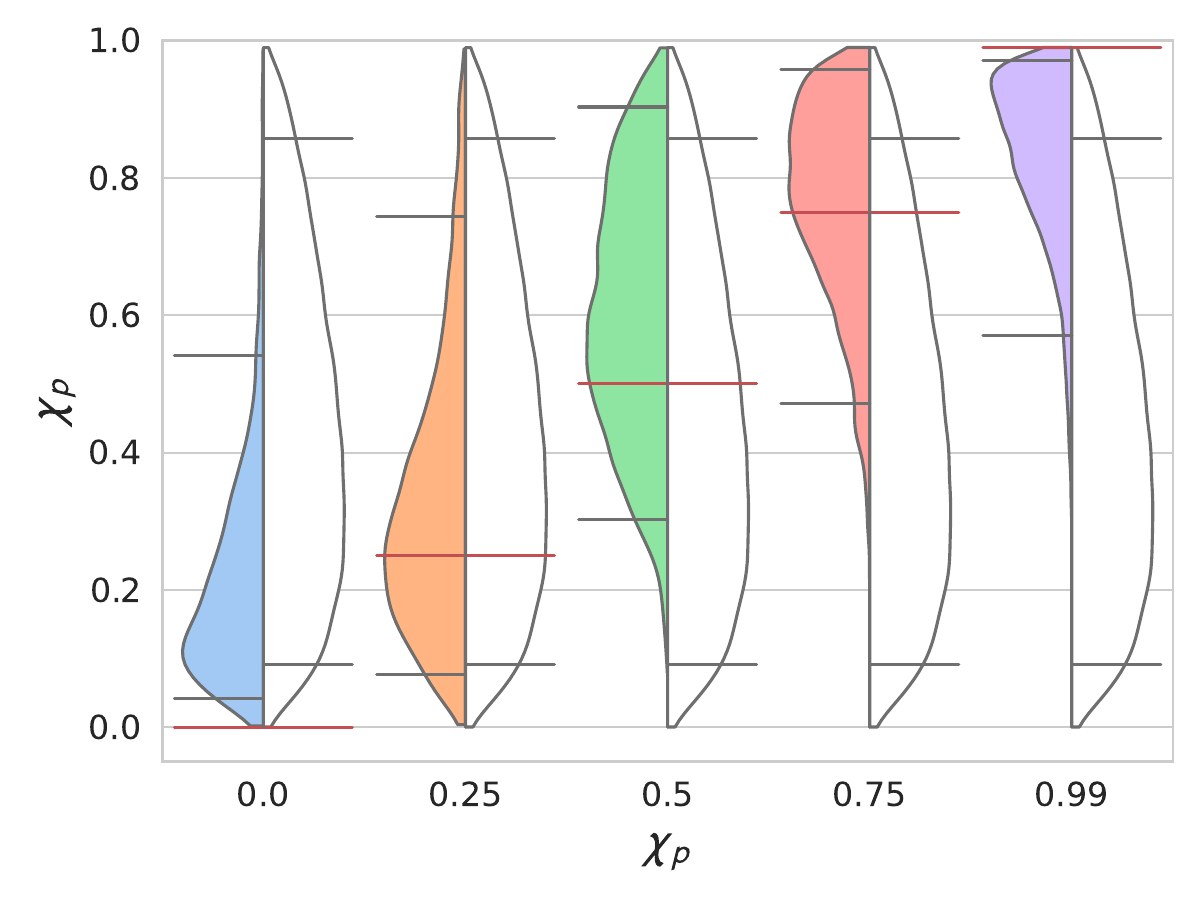}
	\includegraphics[scale=0.4]{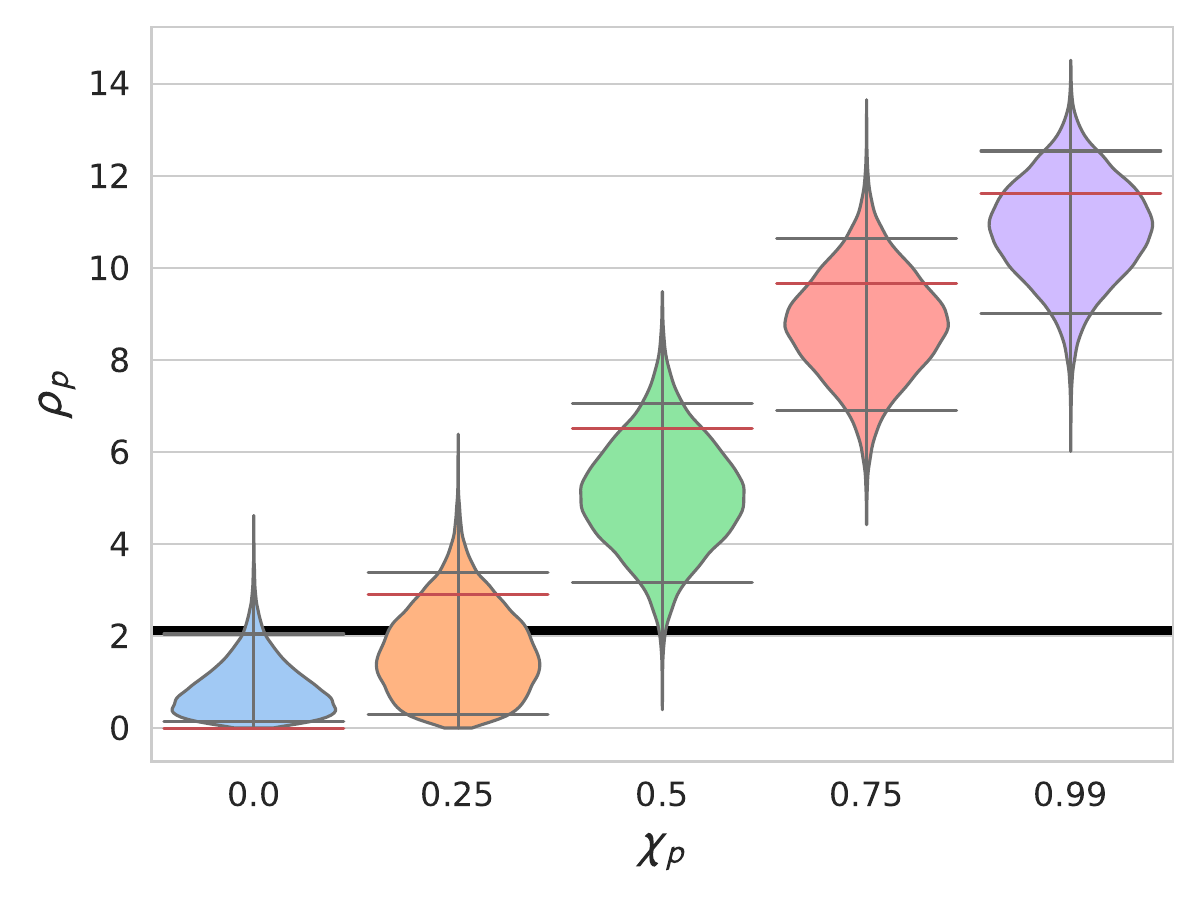}
	\caption{Violin plots showing the recovered posterior distributions distributions for $\chi_p$ compared to its prior (left) and $\rho_p$ (right). Distributions are plotted for varying $\chi_{p}$. Parameters other than $\chi_{p}$ match the ``standard injection'' (see Table~\ref{tab:standard_run})}
	\label{fig:chi_p_vs_chi_p}
\end{figure*}

We now look at the effect of varying the amount of precession in the system,  varying $\chi_p$
from 0 to 1 in steps of 0.25. At $\chi_{p} = 1$ we have maximal spin, all in the plane of the binary.
The inferred values of precessing spin and precession \ac{snr} are shown in Fig.~\ref{fig:chi_p_vs_chi_p}.
We observe, as expected, that increasing the in-plane spin leads to an increase in
the magnitude of precession effects observable in the system. With zero precessing spin, there
is no evidence for precession in the system; the recovered $\chi_p$ is consistent with zero%
\footnote{We do not expect the $\chi_p$ posterior to contain $\chi_p = 0$ as there is no prior support there, however the posterior is relatively well constrained at low precession.}.
Similarly, there is no support for significant precession SNR, with $\rho_p$ constrained near zero.
As $\chi_p$ increases, the amount of precession in the system grows and the measurement of $\chi_p$ becomes both more accurate and more precise.
Fig.~\ref{fig:chi_p_vs_chi_p} shows the relationship between $\rho_p$ and $\chi_p$, and a larger value for $\rho_p$ enables a better measurement for $\chi_p$.

\begin{figure*}[t]
	\centering
	\includegraphics[scale=0.4]{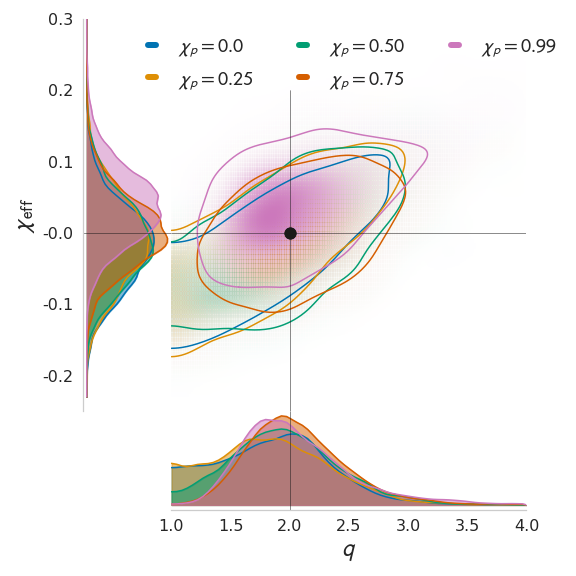}
	\includegraphics[scale=0.4]{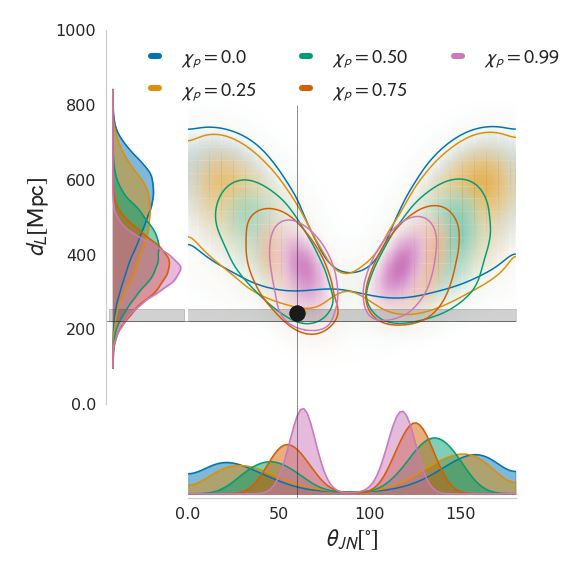}
	\caption{Two dimensional posteriors for (left) mass ratio and aligned spin, $\chi_{\mathrm{eff}}$, (right) binary orientation and distance. Contours show the 90\% confidence interval. Bounded two-dimensional KDEs are used for estimating the joint probability density. The black circle with corresponding horizontal and vertical lines indicates the simulated values. For the simulated distance, a solid horizontal band indicates the maximum and minimum simulated values.
	}
	\label{fig:chi_p_insight_contours}
\end{figure*}

Fig.~\ref{fig:chi_p_insight_contours} shows how the inferred mass ratio--aligned spin and
distance--orientation contours change as the magnitude of the in-plane spins change.  When
there is no observable precession in the system, there is a clear degeneracy in both cases.
However, as precession effects become stronger the degeneracy between both pairs of
parameters is broken.  If $\rho_p$ is small then this can be explained by both a small amount of
precession observed at almost any inclination angle, or a large $\chi_p$ observed close to face on, as seen in
Fig.~\ref{fig:rho_p_chi_p}.  Since precession effects are not strong enough
to provide an accurate measurement of the orientation, the degeneracy between distance and $\theta_{JN}$ persists.
When $\rho_p$ clearly excludes small values, there is \emph{no support} for close to face-on signals, allowing a more precise measurement
of the inclination angle $\theta_{JN}$, breaking the degeneracy with distance.

Stronger precession also allows for improved measurement of the mass ratio.  The opening angle $\beta$, and consequently the
precession parameter $\bar{b}$, increases as the mass-ratio is increased, as can be seen
from Eq.~(\ref{eq:opening_angle}).
Thus, when strong precession effects are observed,
the signal is inconsistent with an equal mass system.  In addition, the difference in
frequency between the two leading precession harmonics depends upon the mass-ratio \cite{Fairhurst:2019_2harm}, and
this may also improve our measurement of $q$. This can also be seen from the precession dynamics,
where the precession rate of $L$ around $J$, $\dot{\alpha}$, depends the mass ratio, and
the number of observable precession cycles corresponds to improved accuracy in the measurement of the
mass ratio~\cite{OShaughnessy:2014shr}.

As $\chi_p$ is increased, the peak of the recovered $\rho_p$ distribution is closer to
the simulated value.  This is likely due to a better
measurement of the binary orientation as shown in Fig. \ref{fig:chi_p_insight_contours}.


\subsection{Inclination}
\label{ssec:inclination}
\begin{figure*}[t]
	\centering
	\includegraphics[scale=0.4]{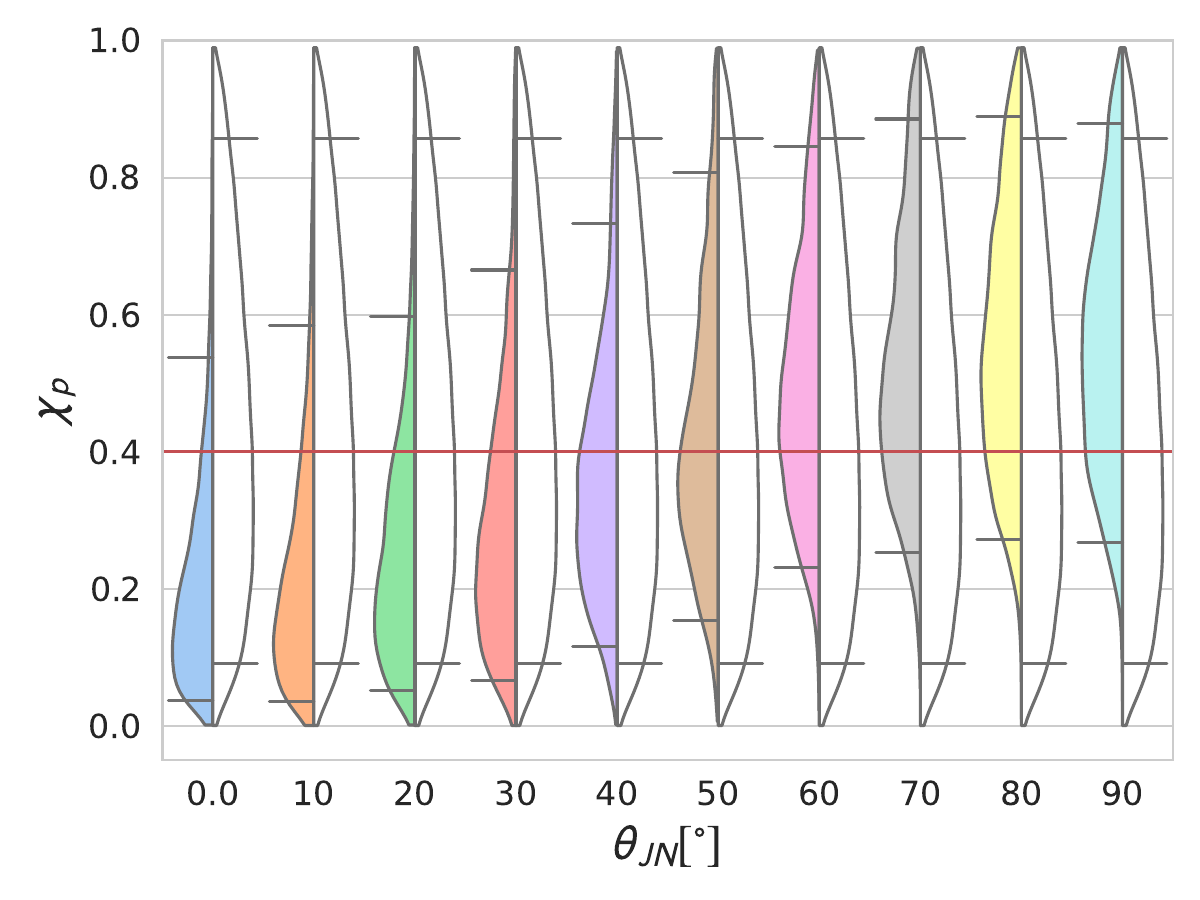}
	\includegraphics[scale=0.4]{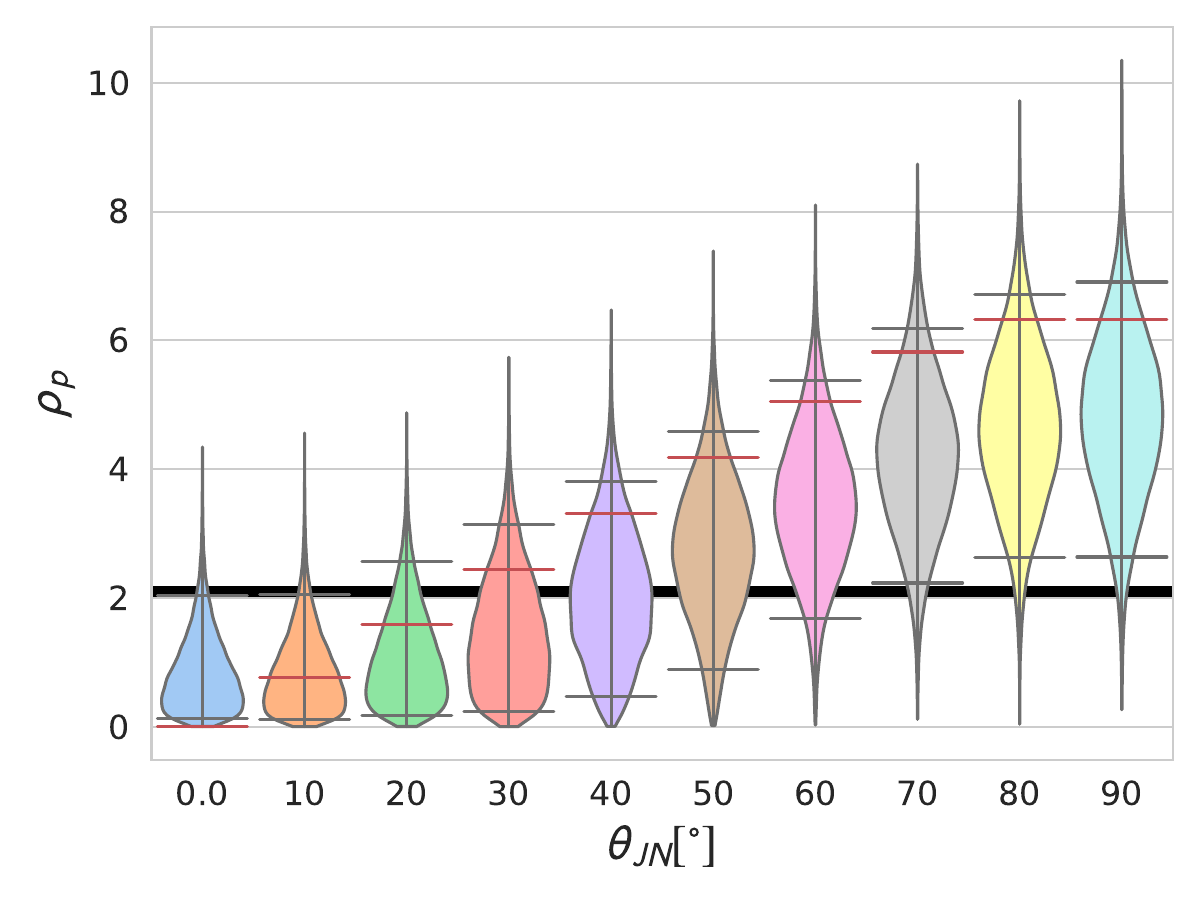}
	  \caption{Violin plots showing the recovered posterior distributions distributions for $\chi_p$ compared to its prior (left) and $\rho_p$ (right). Distributions are plotted for varying  $\theta_{JN}$. Parameters other than $\theta_{JN}$ match the ``standard injection'' (see Table~\ref{tab:standard_run})}
	\label{chi_p_vs_thetaJN}
\end{figure*}

It is well known that the inclination angle will
affect our ability to measure precession, as outlined in the discussion in Sec.~\ref{sec:precession}.
In particular, from Eq.~(\ref{eq:zeta}) we see that in the two-harmonic approximation the second harmonic vanishes when $\theta_{\rm{JN}} = 0^{\circ}$ or $180^{\circ}.$
In this section we consider the effect of changing the orientation of our standard configuration, which allows
us to quantify how it will manifest in realistic LIGO-Virgo signals. A related study has looked at the effect at higher
mass ratios~\cite{Pratten:2020igi}.

\begin{figure}[t]
	\centering
	\includegraphics[width=0.47\textwidth]{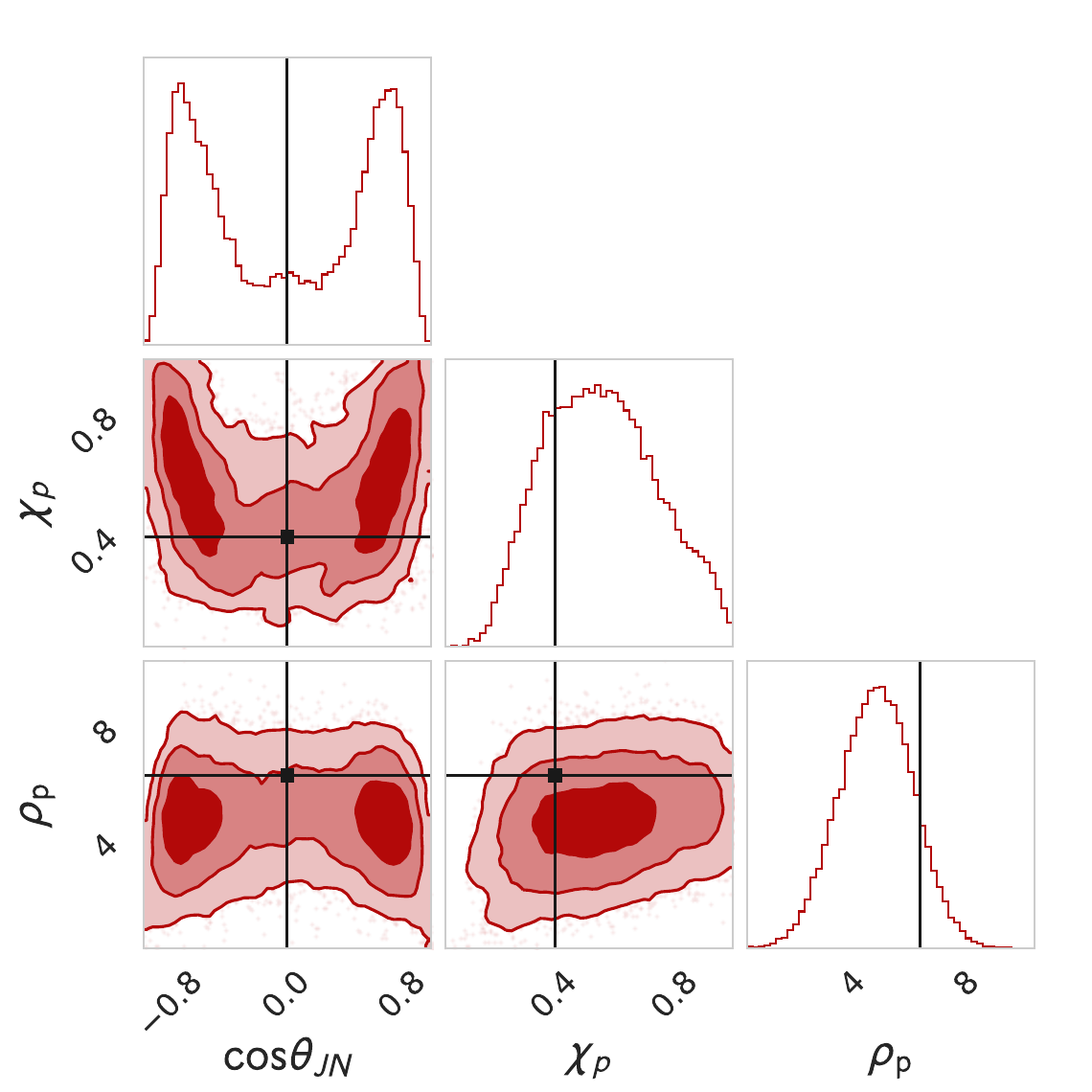}
	  \caption{A corner plot showing the recovered values of binary orientation $\theta_{JN}$, precessing spin $\chi_{p}$ and precession \ac{snr} $\rho_{p}$ for a system simulated at edge on. Shading shows the $1\sigma$, $3\sigma$ and $5\sigma$ confidence intervals. Black dots show the simulated values, We see the strong correlation between $\theta_{JN}$ and $\chi_{p}$ reflecting the measurement of a certain $\rho_{p}$  }
	\label{fig:inc_rho_p_chi_p_contour}
\end{figure}

The effect of varying $\theta_{\rm{JN}}$ is shown in Fig.~\ref{chi_p_vs_thetaJN}.
For binaries where the total angular momentum is nearly aligned with the line of sight, 
precession effects are not observable, as is clear from both the $\rho_p $ and $\chi_p$ posteriors.  It is not until $\theta_{JN} \ge 40\degree $ that we begin to be able to measure precession.  Although the accuracy of the measurement clearly improves as we increase $\theta_{JN}$, the uncertainty in the measurement of $\chi_{p}$ remains large and even at $\theta_{JN} = 90\degree$ the posterior is very broad.  This can be understood by considering the degeneracies shown in Fig.~\ref{fig:rho_p_chi_p} for the standard signal and in Fig.~\ref{fig:inc_rho_p_chi_p_contour} for the $\theta_{JN} = 90^{\circ}$ signal.  In both cases, the measured quantity, $\rho_p$, is relatively well constrained but neither the binary orientation nor $\chi_p$ are accurately measured.  The observed precession is consistent with both a highly inclined system with lower precessing spin (i.e., low $\chi_p$ and large $\theta_{JN}$) or by a less inclined system with higher precessing spin (i.e., high $\chi_p$ and small $\theta_{JN}$).   Both of these will produce similar observable effects in the waveform.

This allows us to explain the measured posterior for $\chi_p$. At low inclination the posterior is consistent with small values of $\chi_p$.  While we are unable to rule out large $\chi_p$, there is limited support as it would require the system to be observed very close to face-on, otherwise precession effects become significant.  At large values of $\theta_{JN}$, when precession is clearly observable in the signal, $\chi_p = 0$ is excluded but the distribution remains broad and extends to $\chi_p = 1$.


\subsection{Mass ratio and aligned spin}
\label{ssec:q_chi}
\begin{figure*}[t]
        \centering
        \includegraphics[scale=0.4]{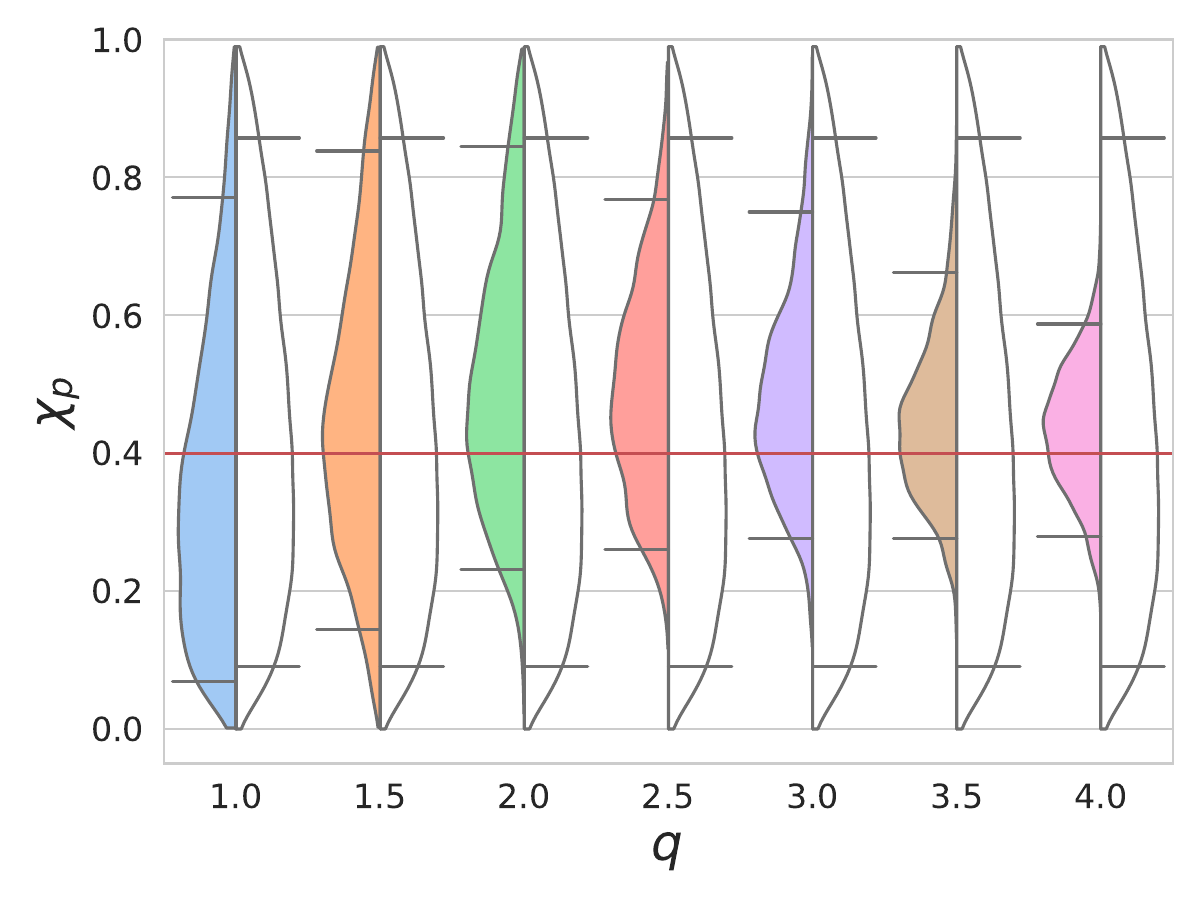}
        \includegraphics[scale=0.4]{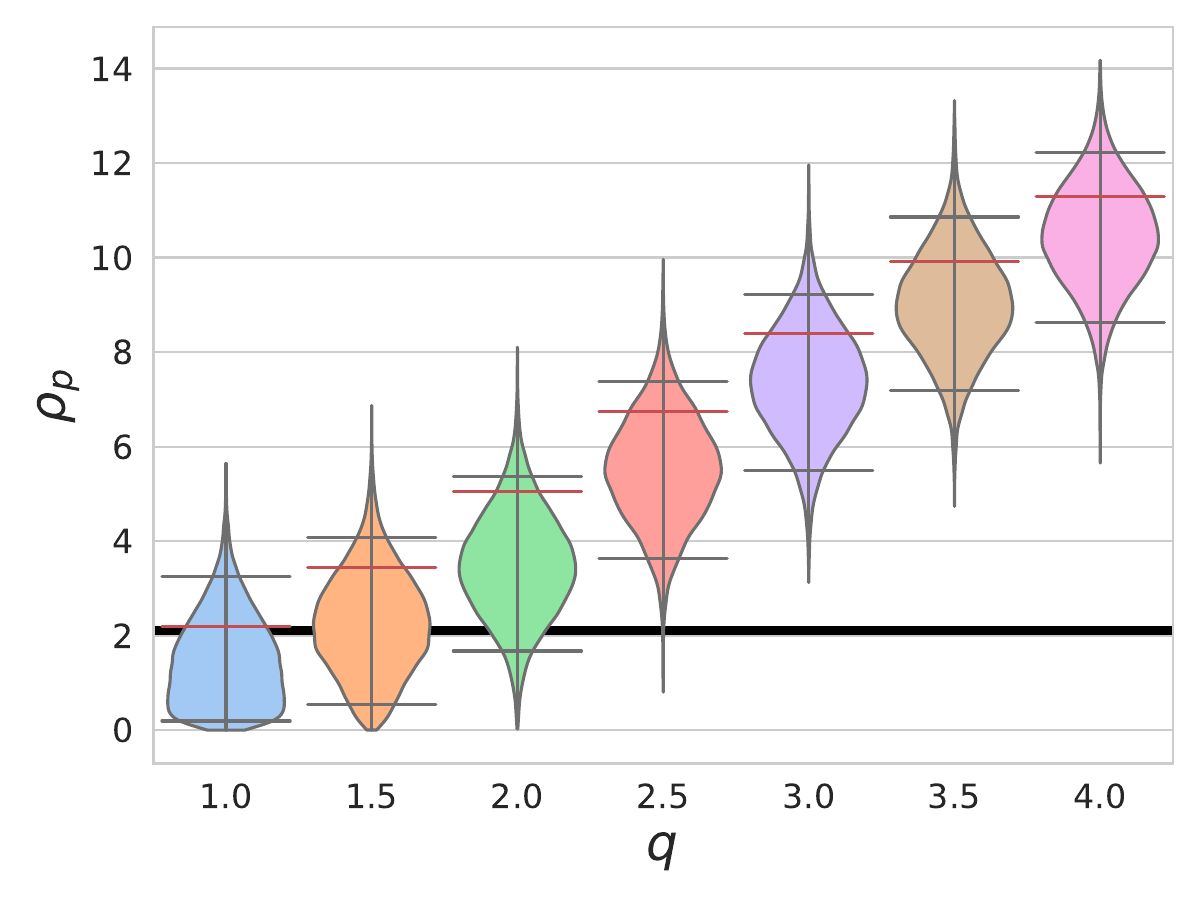}
        \caption{ Violin plots showing the recovered posterior distributions distributions for $\chi_p$ compared to its prior (left) and $\rho_p$ (right). Distributions are plotted for varying mass ratio. Parameters other than the mass ratio of the signal match the ``standard injection'' (see Table~\ref{tab:standard_run}).}
        \label{chi_p_vs_mass_ratio}
\end{figure*}

\begin{figure*}[t]
	\centering
	\includegraphics[scale=0.4]{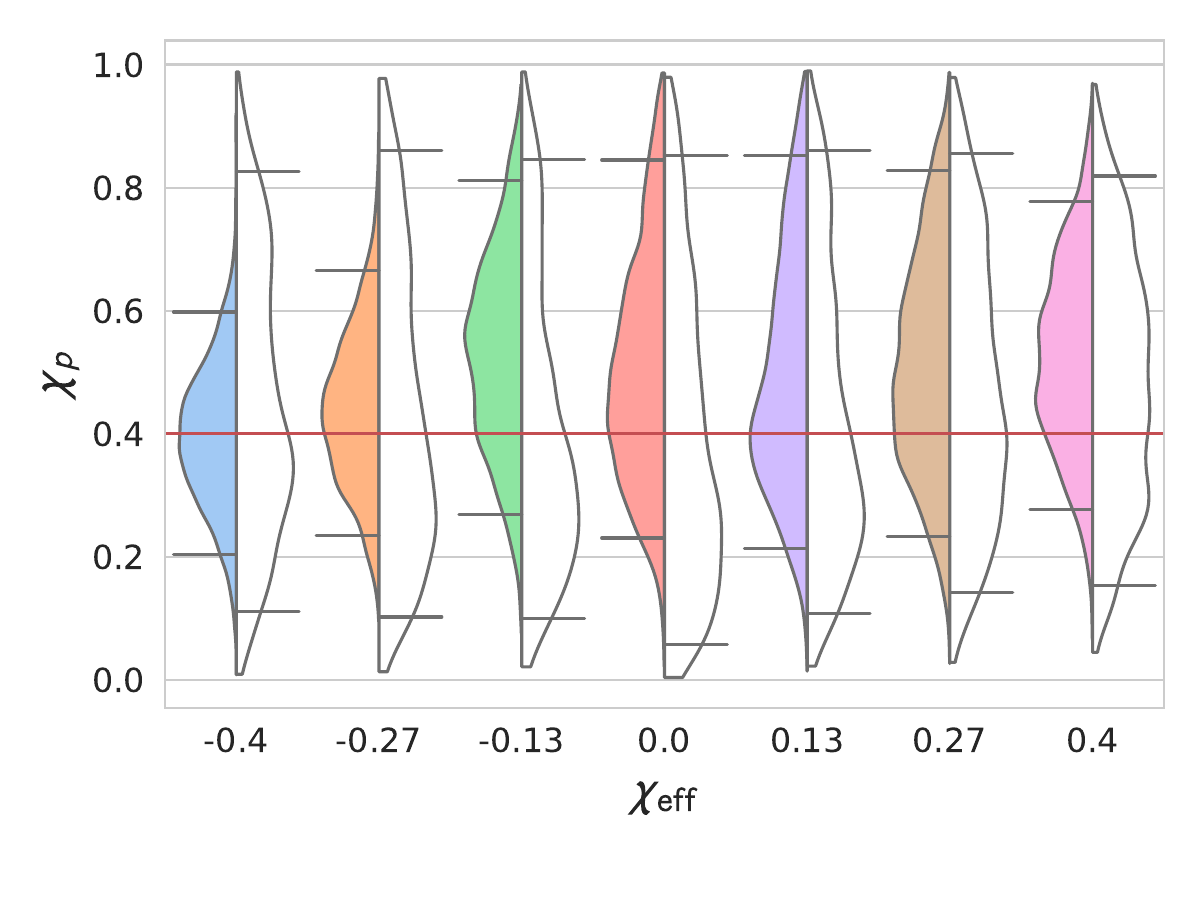}
	\includegraphics[scale=0.4]{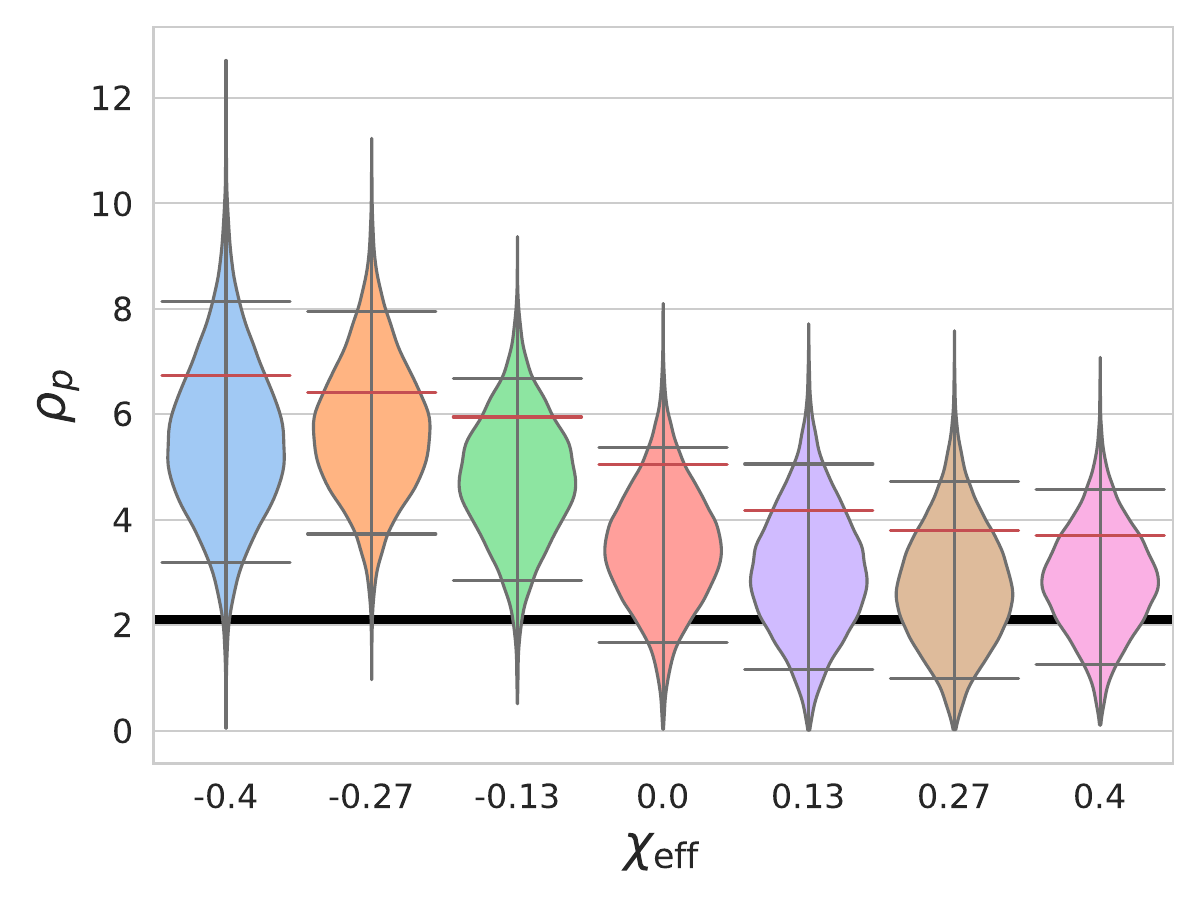}
	\caption{ Violin plots showing the recovered posterior distributions distributions for $\chi_p$ compared to its prior conditioned on the $\chi_{\mathrm{eff}}$ and mass ratio posterior distributions (left) and $\rho_p$ (right). Distributions are plotted for varying $\chi_{\mathrm{eff}}$. Parameters other than the $\chi_{\mathrm{eff}}$ of the signal match the ``standard injection'' (see Table~\ref{tab:standard_run}).}
	\label{chi_p_vs_chi_eff}
\end{figure*}


Fig.~\ref{chi_p_vs_mass_ratio} shows how the inferred precessing spin and precession \ac{snr} varies with the mass ratio of the system.  As expected
from the general considerations presented in Sec.~\ref{sec:precession}, as the mass ratio increases, an in-plane spin on the larger \ac{bh}
leads to a larger opening angle and more significant precession effects.
For near equal-mass systems
($q \lesssim 1.5$), the inferred $\chi_{p}$ posterior distribution resembles its prior, and there is not significant power in precession, as shown by the value of $\rho_{p}$.
As the mass ratio increases, the inferred power in precession also increases and for $q \gtrsim 2.5$, the 90\% credible interval of the
inferred $\rho_{p}$ distribution is entirely above $\rho_{p} = 2.1$.  At this stage, precession is clearly identified and $\chi_{p} \approx 0$
is clearly excluded.  In addition, the maximum value of $\chi_p$ is also bounded away from maximal.

Fig.~\ref{chi_p_vs_chi_eff} shows how varying $\chi_{\mathrm{eff}}$ affects our ability to measure precession.
A system with a large negative $\chi_{\mathrm{eff}}$ results in a larger
opening angle compared to an equivalent system with a large positive $\chi_{\mathrm{eff}}$.  Thus, based upon
Eq.~(\ref{eq:opening_angle}), we expect the observable impact of precession to be greater for negative values of
$\chi_{\mathrm{eff}}$ and smaller for positive values.  The results are consistent with this expectation, in that the precession SNR
decreases with increasing $\chi_{\mathrm{eff}}$ and the width of the recovered $\chi_p$ distribution increases.
However, for the $\chi_{\mathrm{eff}} = 0.4$ analysis, we find that the range of $\chi_{p}$ is restricted, with
both $\chi_{p} = 0$ and $\chi_{p} = 1$ excluded.  This is \textit{not} due to the measurement of precession, but
is actually due to the measured non-zero aligned-spin component.

\begin{figure}[t]
	\centering
	\includegraphics[width=0.47\textwidth]{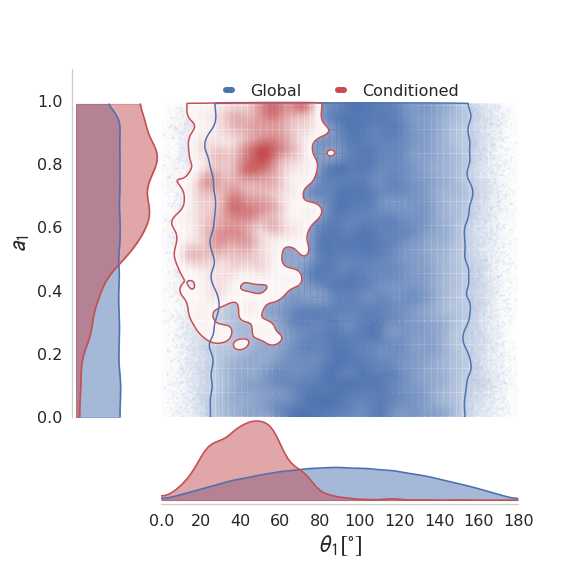}
	\caption{2d contours showing the prior 90\% credible interval over the primary spin magnitude and
	spin direction parameter space. Blue shows the global prior and red shows the global prior
	conditioned on the $\chi_{\mathrm{eff}} = 0.4$ mass ratio and $\chi_{\mathrm{eff}}$ posterior
	distributions}
	\label{fig:conditioned_prior}
\end{figure}

A non-zero measurement of $\chi_{\mathrm{eff}}$ forces $\chi_{p} < 1$ as the primary and
secondary spin magnitudes must be less than unity. For example, in the $\chi_{\mathrm{eff}} = 0.4$ analysis, we measure
$\chi_{\mathrm{eff}} = 0.38^{+0.07}_{-0.07}$. Under the single spin assumption, this limits
$\chi_{p} < 0.95$. Similarly, since we are using prior distributions that are uniform in spin magnitude
and orientation, the observation of a large aligned spin component leads to greater support for a large in-plane
spin component.  This is shown in Fig.~\ref{fig:conditioned_prior}, where we plot both the uninformed prior on the
primary spin as well as the prior conditioned on $\chi_{\mathrm{eff}} = 0.4$, which removes all support for $\chi_p \approx 0$.

The $\chi_{p}$ measurement for the $\chi_{\mathrm{eff}}=0.27$ and $0.4$ analyses are similar to the conditional prior
but do restrict the lower $\chi_{p}$ bound beyond prior effects.  Although the distribution for $\rho_p$ does extend to zero,
it still peaks above $\rho_p = 2.1$ indicating some evidence, although not particularly strong, for precession.

As we vary the mass ratio and aligned spin, the length of the waveform will change.  In particular, the
aligned spin and high mass ratio configurations produce longer waveforms than those with anti-aligned spins and
equal masses~\citep{Campanelli:2006uy}.  In principle, this will impact the measurability of precession, as
longer waveforms allow for a greater number of precession cycles in the detectors' sensitive band.
For very short signals, with less than one precession cycle in band, the two leading harmonics are no longer orthogonal
(or even approximately so), which make it more challenging to unambiguously identify the second harmonic.  This is not
an issue for the signals considered here, but does become important when we vary the mass of the binary in Section \ref{ssec:total_mass}.
With a greater number of precession cycles, we will also be able to more accurately measure the precession frequency (the frequency
difference between the harmonics), which may improve the measurement of mass ratio \cite{OShaughnessy:2014shr}.
However, it is still the precession \ac{snr} that determines the observability of precession.
Finally, we note that changing the mass ratio and aligned spin will change the overall amplitude of the waveform.
Since our study is performed at a \textit{fixed SNR}, this simply leads to the signals being placed at a larger or smaller distance and
therefore doesn't impact the results presented here.


\subsection{Total mass}
\label{ssec:total_mass}

\begin{figure*}[t]
  \centering
	\includegraphics[scale=0.4]{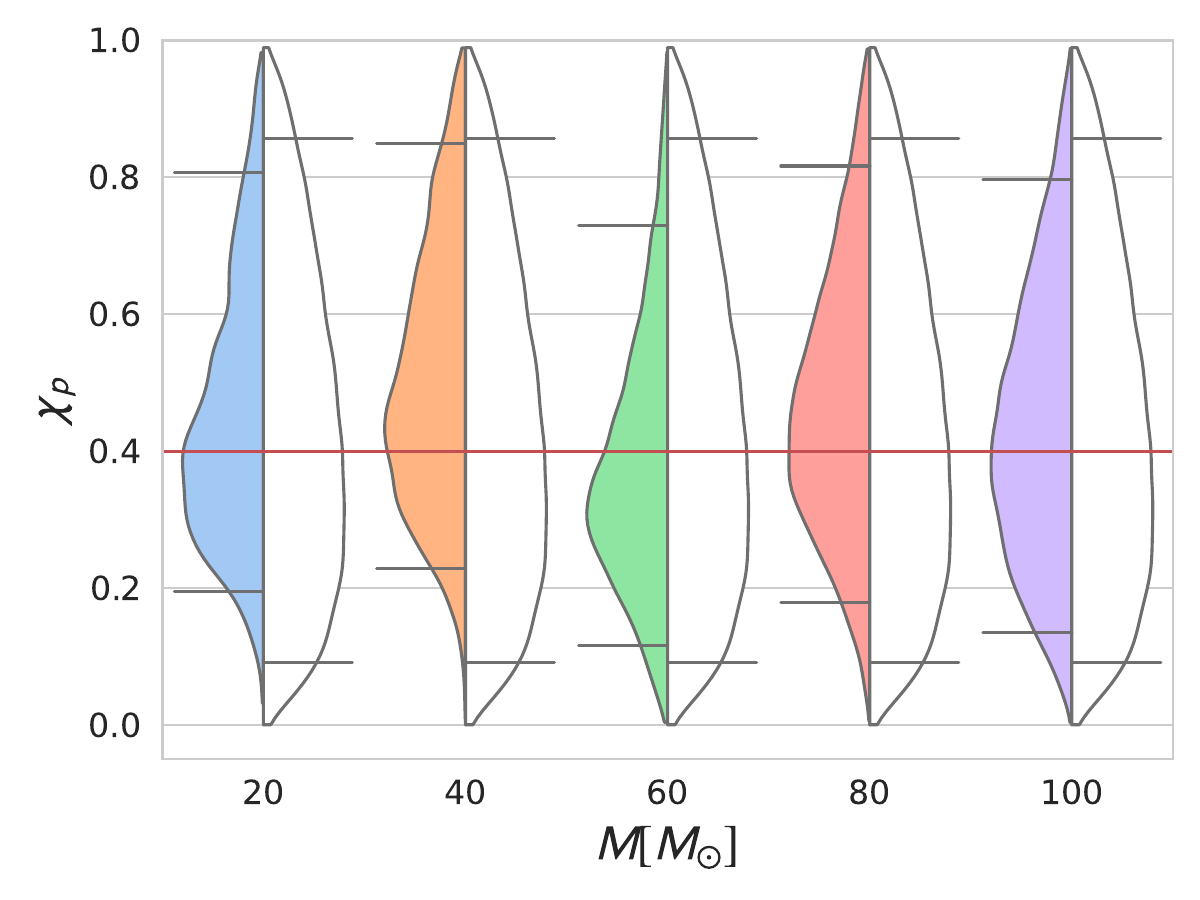}
	\includegraphics[scale=0.4]{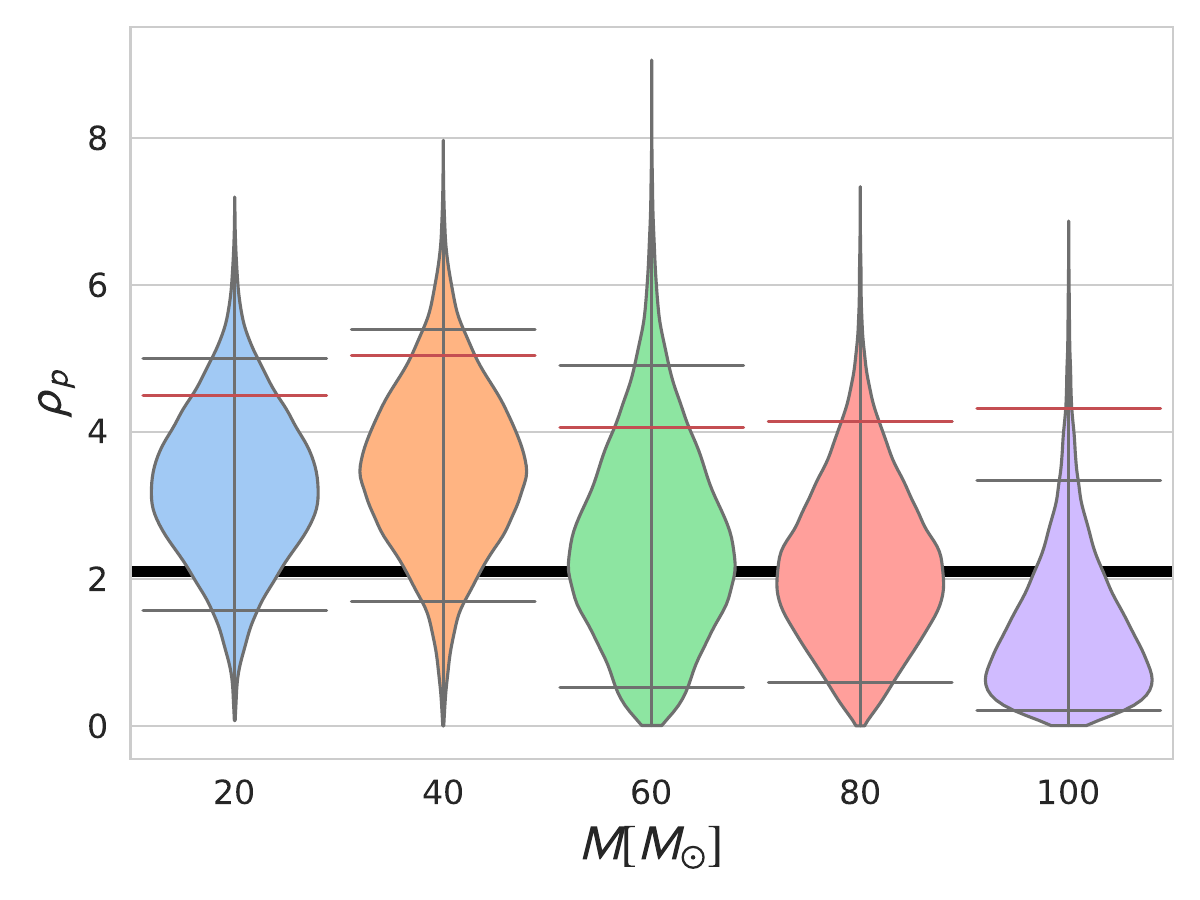}
    \caption{Violin plots showing the recovered posterior distributions distributions for $\chi_p$ compared to its prior (left) and $\rho_p$ (right). Distributions are plotted for varying total mass. Parameters other than the total mass of the signal match the ``standard injection'' (see Table~\ref{tab:standard_run})}
  \label{chi_p_vs_m}
\end{figure*}

We now vary the total mass of the system, keeping all other parameters including mass ratio fixed, in steps of
$20\,M_\odot$.  As before, we keep the \ac{snr} of the system constant at 20, so the higher mass systems are
generated at a greater distance.  The inferred distributions for $\chi_p$ and $\rho_p$ are shown in Fig.~\ref{chi_p_vs_m}.

As the total mass of the source increases, the length of the waveform decreases, as does the
number of precession cycles, with the number scaling approximately inversely to the total mass (see Eq.~(45) of \citep{Apostolatos:1994mx}).
From the two-harmonic perspective, a small number of precession cycles leads to a large overlap between the harmonics.
Specifically, for the $M=100 M_{\odot}$ system
the overlap between the normalised harmonics is $\langle \hat{h}_0| \hat{h}_1 \rangle = 0.77$, where $\hat{h} = h/|h|$
and the inner product is defined in Eq.~(\ref{eq:inner_prod}).  At $M=20M_{\odot}$, the harmonics are close to orthogonal
with $\langle\hat{h}_0| \hat{h}_1 \rangle = 0.15$.  The opening angle doesn't change significantly, with $\bar{b} = 0.14$ at $M=20 M_{\odot}$
and $\bar{b} = 0.21$ at $M=100 M_{\odot}$.

At lower masses, $ M \le 40 M_{\odot}$, while the precessing spin is not tightly
constrained, it is clearly restricted to be non-zero and the precession \ac{snr} has
essentially no support for $\rho_p = 0$.  For the $60 M_{\odot}$ and $80 M_{\odot}$
mergers, the precessing spin is still peaked close to the simulated value while
$\rho_p$ peaks above 2.1 showing evidence for observable precession, although both $\rho_p$ and
$\chi_p$ distributions do extend to zero.

For the high-mass system,
$M = 100 M_{\odot}$, the $\chi_p$ posterior more closely matches the prior and we are unable
to exclude $\chi_p = 0$. The inferred $\rho_{p}$ distribution peaks close to
zero, and is consistent with no precession, even though the precession \ac{snr} in
the simulated signal is
similar to the lower mass signals.  This is likely due to the breakdown of the
two-harmonic approximation for this short signal.  In particular, for a high-mass system, the power
orthogonal to the leading harmonic will depend sensitively upon the initial
precession phase $\phi_{JL}$.  The fact that the recovered value of $\rho_p$ is
inconsistent with the simulated value may be due to this fact: the value of $\phi_{JL} = 45^{\circ}$
used in the simulation leads to maximal observable precession.  Across the
full parameter space there are very few configurations with significant precession, so
this observation is dis-favoured by our priors.  We explore the prior effects such as this in detail in Sec.~\ref{ssec:predict_np}.


\subsection{Polarization}
\begin{figure*}[t]
	\centering
	\includegraphics[scale=0.4]{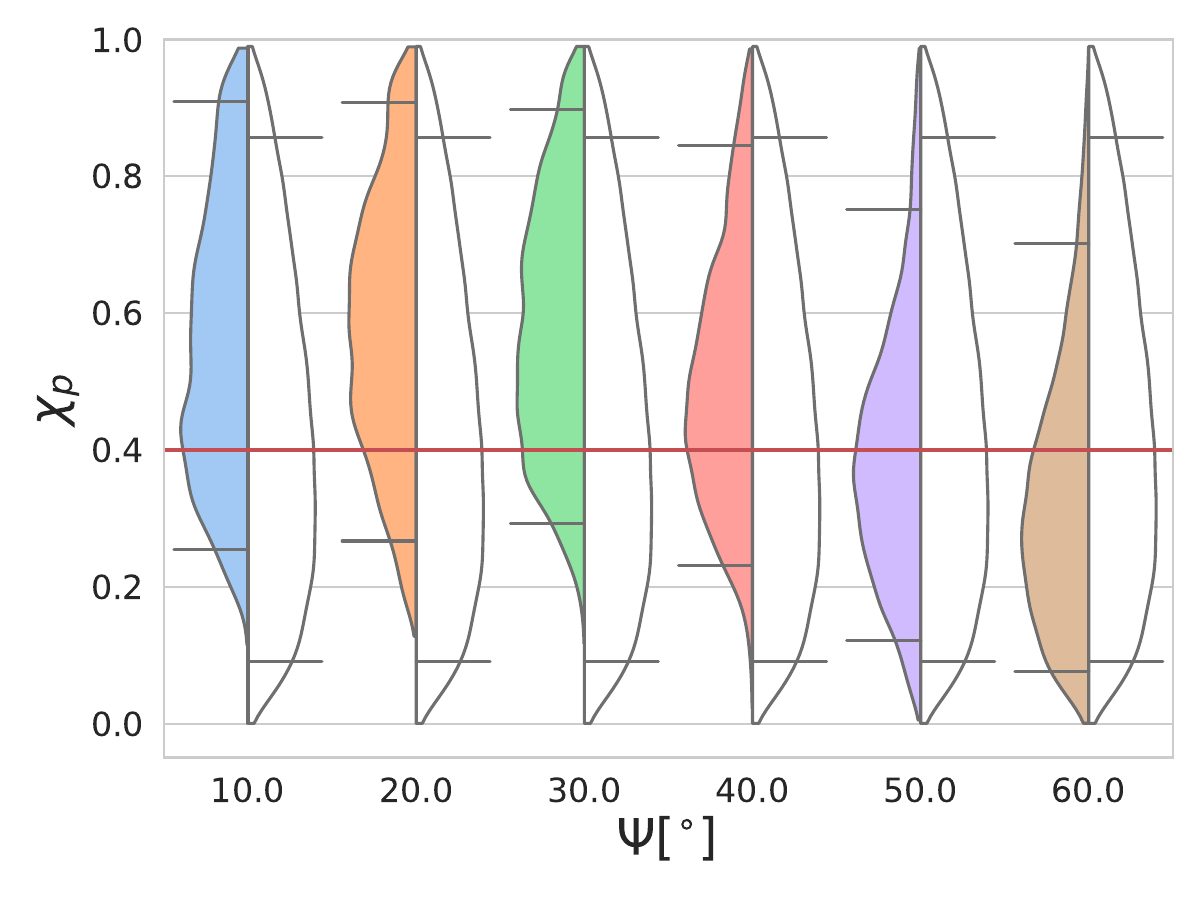}
	\includegraphics[scale=0.4]{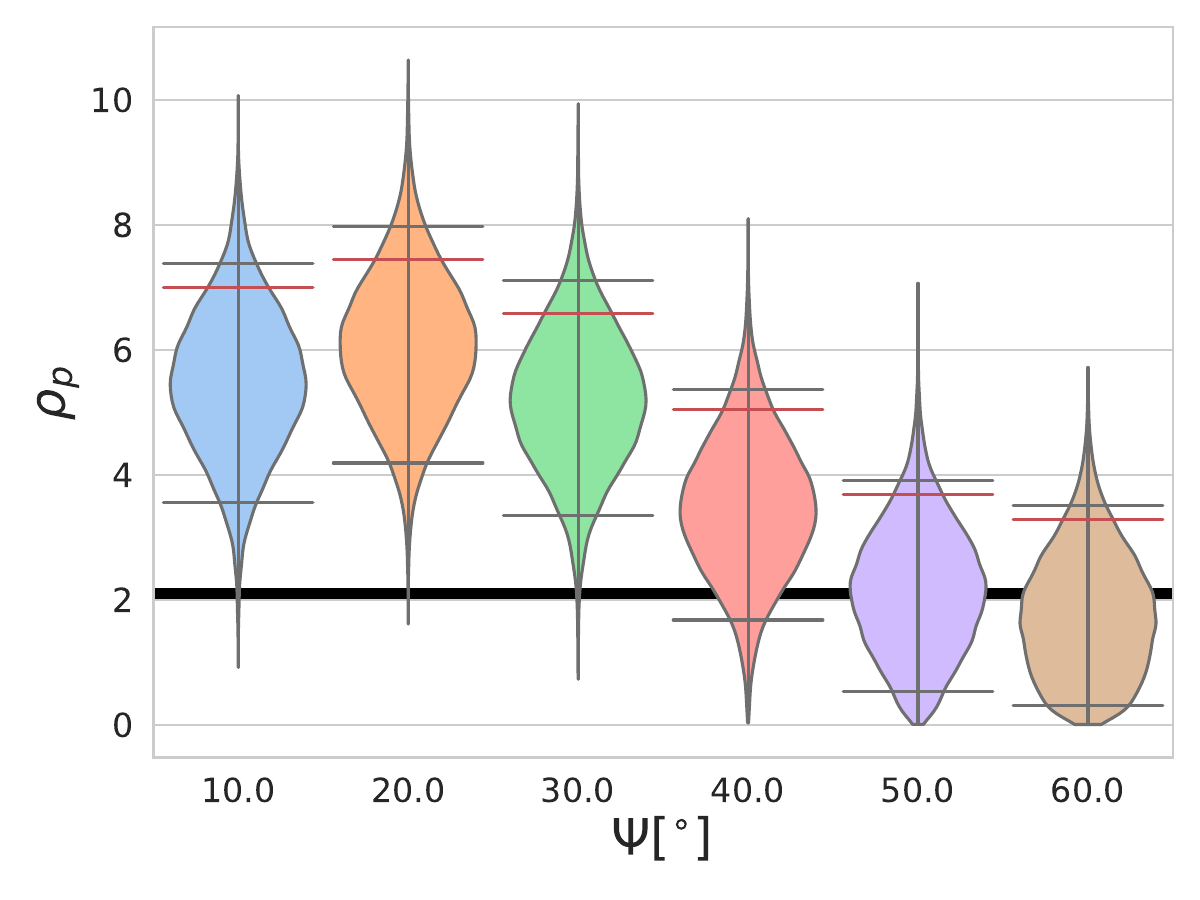}
	  \caption{Violin plots showing the recovered posterior distributions distributions for $\chi_p$ compared to its prior (left) and $\rho_p$ (right). Distributions are plotted for varying  $\psi_J$. Parameters other than $\psi_J$ match the ``standard injection'' (see Table~\ref{tab:standard_run})}
	\label{fig:chi_p_vs_psi}
\end{figure*}

The effect of changing the relative sensitivity to the two GW polarizations is clear from Eq.~(\ref{eq:zeta}).  Recalling that $\bar{b} = 0.11$ and $\theta_{JN} = 60^{\circ}$, we can express $\zeta$ (the ratio of the amplitudes of the two harmonics) as
\begin{equation}\label{eq:zeta_pol}
	|\zeta| = 0.15  \left|\frac{
	F_{+} + 2 i F_{\times} }{1 F_{+} + 0.8 i F_{\times} }\right|  \, , \nonumber
\end{equation}
Thus, $\zeta$, and consequently the imprint of precession on the waveform, will be maximized when the detector network is primarily sensitive to the $\times$ polarization and minimized when the network is sensitive to the $+$ polarization.  We can investigate this by varying the polarization angle of the simulated signal, in steps of $10\degree$ from the ``standard'' value of $40\degree$.  At $\psi = 40\degree$, the sensitivity to the two polarizations is approximately equal, $|F_{\times} |/|F_{+}| = 0.9$.  It is largest for $\psi = 20\degree$ where $|F_{\times}|/|F_{+}| = 25$ and smallest for $\psi=60\degree$ where $|F_{\times}|/|F_{+}| = 0.04$.  This leads to a variation in the precession \ac{snr} from $\rho_{p} \approx 3$ to $\rho_{p} \approx 7$.

In Fig.~\ref{fig:chi_p_vs_psi} we show the recovered posteriors for $\chi_p$ and $\rho_p$ for a set of runs where the precession is varied.
The precession \ac{snr} varies in accordance with expectation --- it is largest at $\psi = 20\degree$, where the median of the posterior is at $\rho_p = 6$ and there is no support for non-precessing systems, and smallest at $60\degree$ where the posterior extends down to $\rho_p = 0$.
The amount of observable precesssion directly impacts the inferred distribution for $\rho_p$.  For the $\psi = 60\degree$ signal, the posterior for $\chi_p$ is consistent with zero, or small in-plane spins, and large values are excluded.  Meanwhile for $\psi = 20\degree$, $\chi_p < 0.1$ is excluded while extremal in-plane spins are consistent with the observation.

It is well known that precession leaves a stronger imprint upon the $\times$ polarization.  However, we are not aware of previous results showing how simply changing the polarization of the system can so dramatically change the observable consequences of precession --- from being barely observable when the observed signal is primarily the $+$ polarization to being strongly observed in $\times$.  Using the two-harmonic approximation, we are able to straightforwardly predict this effect and then verify it with detailed parameter estimation studies.


\subsection{Sky Location}
\label{ssec:sky_loc}
\begin{table}[t]
    \begin{ruledtabular}
        \begin{tabular}{ c | c c c c c | c | }
            Label & $\mathrm{RA} / \mathrm{rad}$ & $\mathrm{DEC} / \mathrm{rad}$ & $\psi/{}^{\circ}$ & $d_{L}/\mathrm{MPc}$ & $\rho_{p}$ & $d_{L}/\mathrm{MPc}$ \\
            \hline
            A & 0.31 & 0.92 & 320 & 370 & 5.02 & $480^{+130}_{-180}$ \\
            B & 0.80 & 1.15 & 345 & 320 & 5.09 & $470^{+140}_{-160}$ \\
            C & 1.31 & 1.22 & 10 & 280 & 5.11 & $450^{+150}_{-160}$ \\
            D & 1.88 & 1.19 & 40 & 220 & 5.05 & $430^{+160}_{-160}$ \\
            E & 6.11 & 0.21 & 40 & 310 & 5.09 & $440^{+150}_{-170}$ \\
            \end{tabular}
        \end{ruledtabular}
    \caption{Table showing the simulated parameters for the sky location set (see Sec.~\ref{ssec:sky_loc}). All other parameters match the ``standard injection'' (see Table~\ref{tab:standard_run}). The recovered luminosity distance (far right column) is also shown.}
    \label{tab:sky_loc_runs}
\end{table}

We performed a series of runs where we altered
the sky location of the signal, keeping the masses
and spins of the components fixed.  We also maintained
the binary orientation $\theta_{\rm JN} = 60\degree$, but
varied the distance and polarization of the source to
ensure that the \ac{snr} remained constant and that the
relative contribution of the $+$ and $\times$ polarizations
was consistent with the standard run.
Furthermore, sky locations were restricted to those for which the relative
time of arrival between the Hanford and Livingston detectors
remains the same (i.e., we were sampling from the nearly
degenerate ring on the sky of constant time delays).
Details of the runs are given in Tab.~\ref{tab:sky_loc_runs}.
\begin{figure*}[t!]
	\centering
	\includegraphics[width=0.68\textwidth]{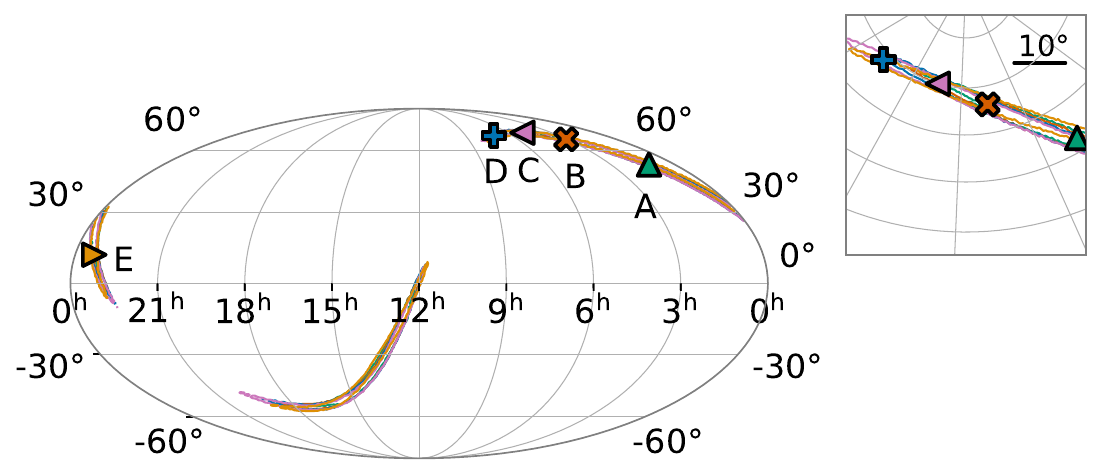}
	\caption{
	Skymap showing the different simulated sky positions, see Table~\ref{tab:sky_loc_runs}. The solid lines
	show the 90\% credible intervals and the markers show the simulated sky position. Their respective colors
	matches their corresponding credible intervals. We vary the distance
	and polarization of the source to ensure that the \ac{snr} remains consistent with the standard
	injection in Table~\ref{tab:standard_run}.
}
	\label{fig:sky_loc}
\end{figure*}

Table~\ref{tab:sky_loc_runs} shows that the inferred luminosity distance
remains approximately constant despite the simulated luminosity distance varying by
almost a factor of two.  In addition, the recovered $\rho_p$ distribution remains
consistent with the ``standard'' injection.
Fig.~\ref{fig:sky_loc} shows that the inferred sky position of the source remains
essentially unchanged, and consistent with locations of the detectors' greatest sensitivity.
We note here that for this study we only considered the two detector LIGO network. 
Including VIRGO would likely have considerably improved the precision of the 
inferred sky location. We do not expect that this would affect any of the inferred
physical parameters or any of the main conclusions in this work.


\section{Relating $\rho_p$ posteriors to Bayes Factors}
\label{sec:bayes_factors}
\begin{figure}[t]
	\centering
	\includegraphics[width=0.49\textwidth]{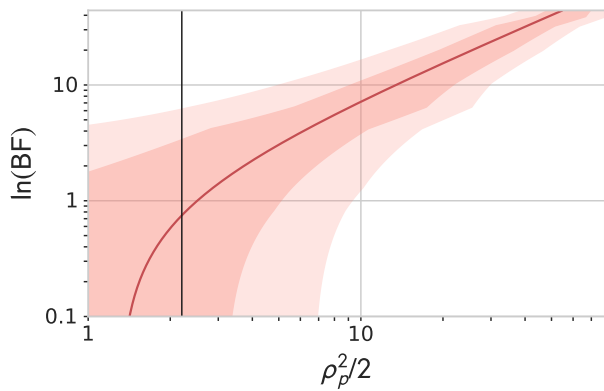}
	\caption{
	Plot comparing the Bayes factor in favour of precession to the inferred $\rho_{p}$
	distribution. Bayes factors were calculated by comparing the evidences for a precessing analysis
	and a non-precessing analysis. The uncertainties on the Bayes factors are calculated by taking the
	$90\%$ confidence interval across multiple {\sc{LALInferenceNest}} chains. The solid line uses
	the median of the $\rho_{p}$ distribution. The shading gives the $1\sigma$ and $2\sigma$ uncertainties
	on the $\rho_{p}$ measurement. The solid black lines shows the $\rho_{p} = 2.1$ threshold.}
	\label{fig:rho_p_vs_bayesfactor}
\end{figure}

An alternative method for identifying evidence for precession can be calculated within the
Bayesian framework. We can calculate the Bayes factor, $\mathcal{B}$, by comparing the
marginalized likelihoods (see Eq.~(\ref{bayesequation})) from two competing
hypotheses (A, B)~\citep{jaynes2003probability},
\begin{equation}
	\ln{\mathcal{B}} = \ln{p(d_{A})} - \ln{p(d_{B})}.
\end{equation}
Bayes factors have thus far been the gold standard
for identifying evidence for precession within the GW community and have been used
extensively in previous works, see e.g., Ref.~\cite{Pratten:2020igi}.

In the same way that Bayes factors can be used to quantify evidence for precession, it is also possible to
quantify the significance of a GW signal by calculating the Bayes factor for signal verses
noise~\citep{Veitch:2009hd}. It has been shown that the log Bayes factor for signal versus noise scales
approximately with $\rho^2$~\citep{cornish2011gravitational}. Here, we investigate
the relationship between the Bayes factor in favour of precession and the precession \ac{snr} $\rho_{p}$.
Both of  these quantities have been used together in recent works when assessing the evidence for
observable precession~\citep{LIGOScientific:2020stg, Abbott:2020khf, Pratten:2020igi}

For a subset of the runs described in Section.~\ref{ssec:inclination}, we reran the analysis
using the aligned-spin waveform model {\texttt{IMRPhenomD}}. Bayes factors in favour
of precession could then be calculated and compared to the derived $\rho_{p}$ posterior
distributions.

Fig.~\ref{fig:rho_p_vs_bayesfactor} shows an approximately linear relationship between the
log Bayes factor ($\ln \mathrm{BF}$) and the square of the precession \ac{snr} ($\rho_{p}^{2}$).
This is expected given that the likelihoods recovered from the precessing waveform model will be
larger than the likelihoods recovered from the aligned-spin waveform model
by a factor of $\exp(\rho_{p}^{2} / 2)$.

The commonly used heuristic when assessing the strength of evidence using Bayes factors is that
 $ 1 \leq \ln\mathrm{BF} \geq 3 $ is marginal evidence and $\ln\mathrm{BF} > 3 $  is strong evidence in favour of a hypothesis.
From the plots above we conclude that if 90\% (50\%) of the $\rho_{p}$ posterior distribution is above the $\rho_{p}=2.1$ threshold, this corresponds to a $\ln\mathrm{BF} \approx 3.5$ ($\ln\mathrm{BF} \approx 0.8$) and is therefore very strong (marginal)
evidence for precession. The posterior distribution on $\rho_{p}$ can therefore
be approximately mapped to the commonly used $\ln\mathrm{BF}$. Assessing the strength of evidence for precession using $\rho_p$ would also reduce the need for additional parameter estimation runs using non-precessing models, which are \emph{necessary} to compute the Bayes factor. This reduction in computational
cost will not be significant for a single event, but for population analyses and large scale PE studies this
alternative metric could be extremely useful.


\section{Predicting the Precession SNR Posterior}
\label{sec:rho_p_predict}
For the majority of simulations presented in this paper, the
distribution for the precession SNR, $\rho_{p}$, has been peaked
significantly below the simulated value, although in nearly every case the
simulated value does lie within the $90\%$ confidence region.  While the
naive expectation is that the recovered posterior will peak at the simulated
value, for complex parameter recovery where there are dependencies and
degeneracies between the different parameters, this is often not the case.
We have already seen that the distance is typically over-estimated in the
simulations we have performed --- this is a well-known effect and arises for
two reasons, first that the network is less sensitive to sources from the chosen
sky location than from other locations consistent with the observed signal (as
discussed in Sec.~\ref{ssec:sky_loc}), and second that the signal
was simulated significantly inclined from face-on, yet preferentially recovered
close to face-on (as discussed in Sec.~\ref{ssec:inclination}).
Similarly, it seems likely that the signals we have
simulated have more significant precession effects (deliberately, as we wish
to understand the observability of precession) than the vast majority of
possible sources.  Thus, our conjecture is that the likelihood peaks at the
simulated value of $\rho_{p}$ but the posterior distribution will be biased to
recover a smaller value owing to the much larger volume
of parameter space consistent with low $\rho_{p}$. To demonstrate this, we
calculate a prior distribution for $\rho_{p}$ which uses the information gleaned from
a non-precessing analysis to take into consideration the much larger volume
of parameter space consistent with low $\rho_{p}$. We then show that when
multiplying the likelihood by the prior, the \emph{predicted} posterior
for $\rho_{p}$ agrees well with the \emph{inferred} posterior from a fully
precessing parameter estimation analysis.

Let us first show that the likelihood peaks at the simulated value of $\rho_{p}$.
The two-harmonic approximation allows us to factorize the likelihood in Eq.~(\ref{likelihood_calculation})
into two terms:
a non-precessing component (dependent on $h_{0}$) $\Lambda_{\mathrm{np}}(\bm{\lambda})$
 and precessing component (dependent on $h_{1}$) $\Lambda_{\mathrm{p}}(\bm{\lambda})$,

\begin{widetext}
\begin{eqnarray} \label{likelihood_calculation}
	p(d|\bm{\lambda})
	&\propto& \exp\left(-\frac{1}{2}
	\langle d - (\mathcal{A}_{0}(\bm{\lambda})h^{0}(\bm{\lambda}) + \mathcal{A}_{1}(\bm{\lambda}) h^{1}(\bm{\lambda})) |
	d - (\mathcal{A}_{0}(\bm{\lambda}) h^{0}(\bm{\lambda}) + \mathcal{A}_{1}(\bm{\lambda}) h^{1}(\bm{\lambda}) \rangle \right) \\
	&\propto& \exp\left(
	  \langle d | \mathcal{A}_{0}(\bm{\lambda}) h^{0}(\bm{\lambda}) \rangle
	 - \frac{|\mathcal{A}_{0}(\bm{\lambda})|^{2}}{2} \langle h^{0}(\bm{\lambda}) | h^{0}(\bm{\lambda}) \rangle\right)
	 \times \exp\left(\langle d | \mathcal{A}_{1}(\bm{\lambda}) h^{1}(\bm{\lambda}) \rangle
	 - \frac{|\mathcal{A}_{1}(\bm{\lambda})|^{2}}{2} \langle h^{1}(\bm{\lambda}) | h^{1}(\bm{\lambda}) \rangle
	 \right) \nonumber \\
	&\propto&\Lambda_{\mathrm{np}}(\bm{\lambda}) \times \Lambda_{\mathrm{p}}(\bm{\lambda}), \nonumber
\end{eqnarray}
\end{widetext}
For simplicity we use the approximations that $\langle h^{0} | h^{1} \rangle = 0$ and
that $h^{0}$ is the dominant harmonic, i.e., that the SNR in the $h^{0}$
harmonic is larger than in $h^{1}$. The calculation proceeds analogously when $h^{1}$ is
dominant, and can be extended to the general case by replacing $h^{1}$ by its projection onto the space orthogonal
to $h^{0}$.

We can re-express the precessing contribution to the likelihood $\Lambda_{p}$ in terms of the precession SNR using Eq.~(\ref{rho_p}).  To do so, we introduce $\hat{\rho}_{p}$ which is the simulated value
of $\rho_{p}$, and $\rho_{p}(\bm{\lambda})$ which is the precession SNR for the set of parameters $\bm{\lambda}$.  Furthermore, we define the simulated phase (as given in Eq.~(\ref{eq:2harm_amps})) of the precession harmonic as $\hat{\phi}_{1}$ and the phase associated with the parameters $\bm{\lambda}$ as $\phi_{1}(\bm{\lambda})$. Following the procedure described in, e.g. Ref.~\cite{mills2020measuring}, we can rewrite the precession likelihood as
\begin{widetext}
\begin{equation}
	\Lambda_{\mathrm{p}}(\rho_{p}, \phi_{1}) \propto  \exp \left( - \tfrac{1}{2} \left( \rho_{p}^{2}(\bm{\lambda})
	- 2\hat{\rho}_{p} \rho_{p}(\bm{\lambda}) \cos (\hat{\phi}_{1} - \phi_{1})
	+\hat{\rho}_{p}^{2} \right) \right)\, .
\end{equation}
\end{widetext}
In general, we have no prior knowledge of the precession phase, so it is natural to assume a uniform prior on $\phi_{1}$. We may then analytically marginalise $\Lambda_{\mathrm{p}}(\rho_{p}, \phi_{1})$ over $\phi_{1}$ to obtain,

\begin{eqnarray}
 	\Lambda_{\mathrm{p}}(\rho_{p}) & \propto & \int_{0}^{2\pi}
  \Lambda_{\mathrm{p}}(\rho_{p}, \phi_{1}) \, p(\phi_{1}) \, d \phi_{1} \\
 	& \propto & I_{0}(\hat{\rho}_{p}\, \rho_{p}) \exp\left( - \frac{\hat{\rho}_{p}^{2} + \rho_{p}^{2}}{2} \right). \nonumber
\end{eqnarray}
We therefore see that the precession likelihood peaks at $\hat{\rho}_{p}$. We may then calculate the
posterior distribution for $\rho_p$ using Bayes' Theorem,

\begin{equation}\label{eq:p_rho_p}
	p(\rho_{p} | d) \propto p(\rho_{p} ) \Lambda_{\mathrm{p}}(\rho_{p}) \, ,
\end{equation}
where $p(\rho_{p})$ is the prior for the precession SNR.

Previously, in Ref.~\cite{Fairhurst:2019_2harm}, we obtained a distribution for $p(\rho_{p} | d) $ by maximising the likelihood over $\mathcal{A}_{1}$. This is equivalent to assuming uniform priors for the real and imaginary components of $\mathcal{A}_{1}$, and leads to a prior $p(\rho_{p}) \propto \rho_p$. It follows from Eq.~\ref{eq:p_rho_p} that this results in a $\chi^2$ distribution with 2 degrees of freedom. Here, we instead use a prior for $\rho_p$ which is informed by the information obtained from a non-precessing analysis, we refer to this as the \emph{informed} prior.  This \emph{informed} prior better represents our prior knowledge about $\rho_p$ before explicitly accounting for precession in our analysis.

The majority of parameters required to calculate the \emph{informed} prior are already given in the non-precessing results.
The two exceptions are the amplitude of the precessing spin $\chi_{p}$ and the
initial precession phase $\phi_{JL}$.  As discussed in Section \ref{ssec:q_chi},
we can obtain a prior for $\chi_{p}$ \textit{conditioned} upon the other
parameters, specifically the mass ratio and aligned spin $\chi_{\mathrm{eff}}$,
and this can be used to generate the informed prior on $\rho_p$.
The initial precession phase is unconstrained by the non-precessing parameter recovery,
this then allows us to assume it to be uniformly distributed.
By calculating the
\emph{predicted} posterior distribution for $\rho_{p}$ based upon a set of non-precessing samples,
we may examine the effect of other measured parameters on the final $\rho_p$ distribution.
For example, if the aligned-spin run favours a
binary that is close to equal mass and an orientation consistent with a face-on
system, then our prior belief will be that the precessing SNR will be low ---
it is only with unequal masses and systems misaligned with the line of sight
that there are significant precession effects in the observed waveform. A prior belief of
$\rho_{p}$ peaking at low values will cause the predicted $\rho_{p}$ to peak at values lower
than the simulated one and consequently so too will the inferred posterior distribution for
$\rho_{p}$ inferred from a full 15-dimensional parameter estimation analysis.

\subsection{Precessing signal}

\label{ssec:predict_prec}

\begin{figure}[t]
	\centering
	\includegraphics[width=0.49\textwidth]{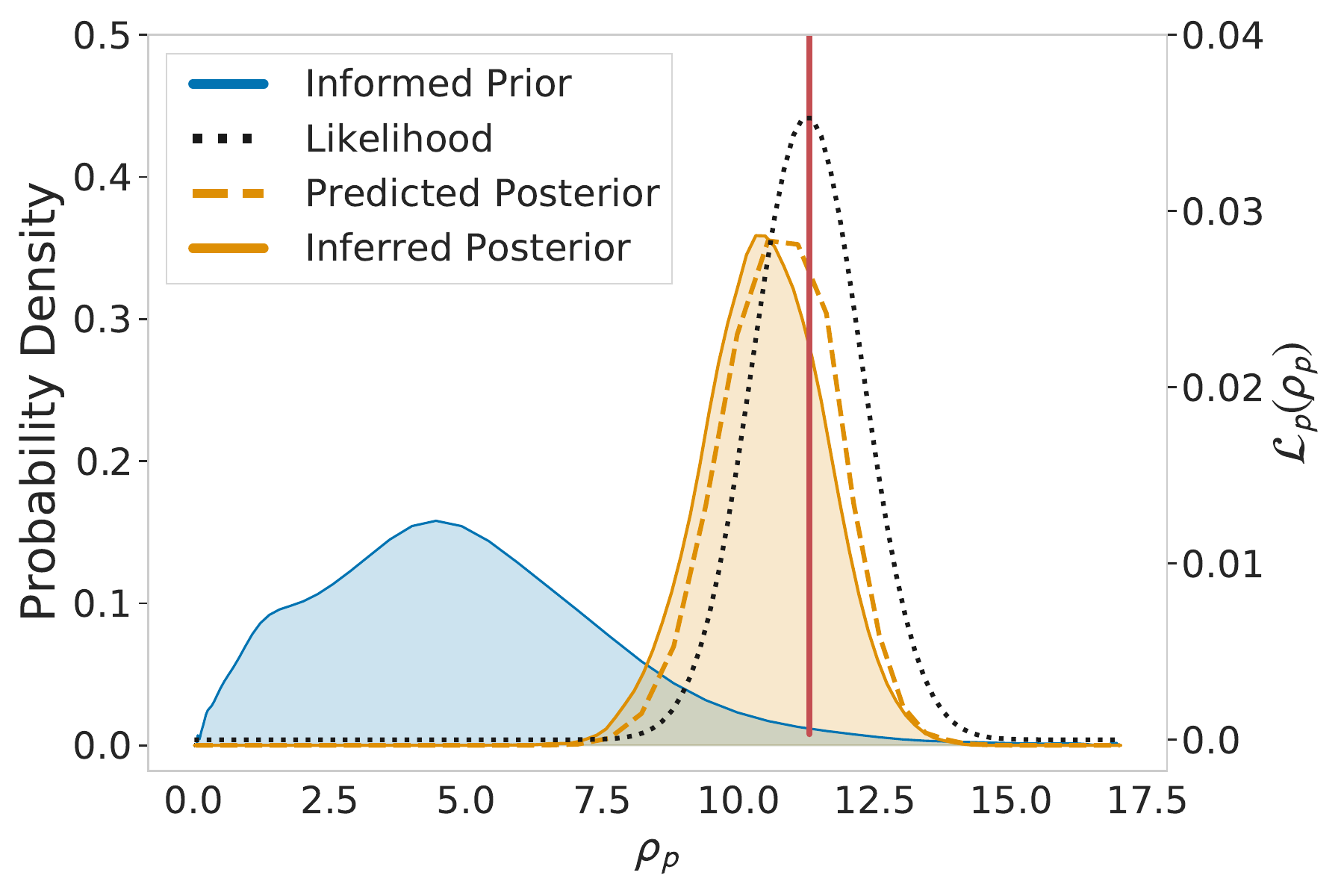}
	\caption{
	The predicted distribution for the precession \ac{snr} $\rho_{p}$ (dashed orange) calculated as the product
	of the precessing contribution to the likelihood (black dotted line) and the informed prior of $\rho_{p}$ (blue) for the $q=4$ simulation presented in Sec.~\ref{ssec:q_chi}. For comparison, we show the inferred $\rho_{p}$ posterior distribution from the full 15 dimensional parameter estimation analysis (solid orange) and $\rho_{p}$ for the injection (red line).
	The informed prior is peaked at low values of $\rho_{p}$ causing the peak of the
	posterior to be smaller than the maximum likelihood value.
	}
	\label{fig:predicted_prior_rho_p}
\end{figure}

\begin{figure}[t]
	\centering
	\includegraphics[width=0.49\textwidth]{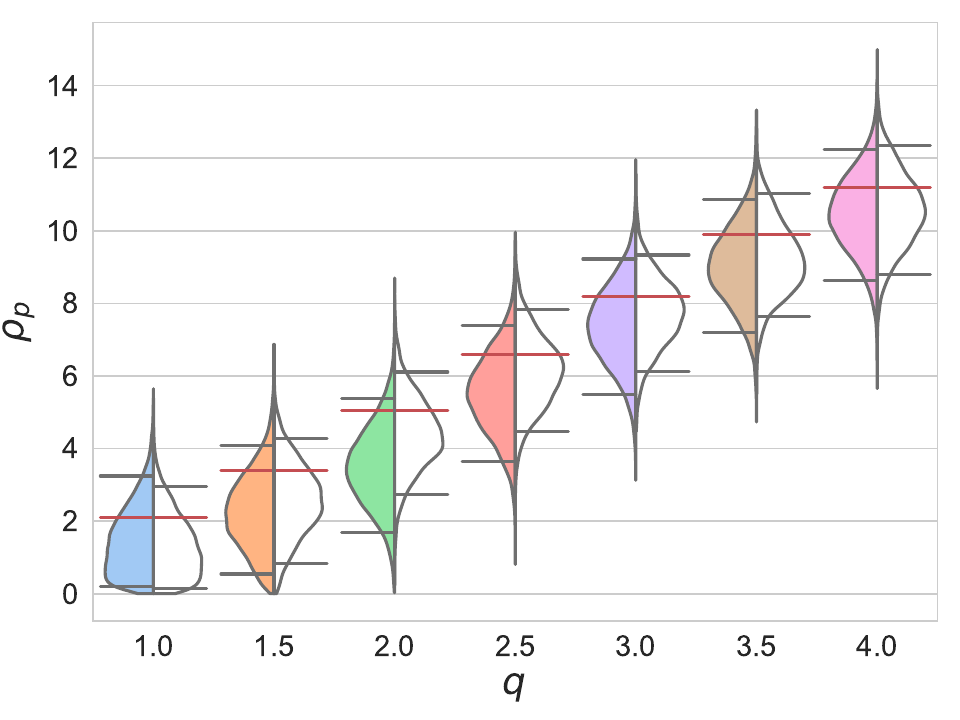}
	\caption{
	Violin plot comparing the observed $\rho_{p}$ distribution (colored) from a precessing analysis,
	and the predicted distribution (white) based on the aligned-spin results and simulated value of $\rho_p$
	for the set of varying mass ratio simulations presented in Sec.~\ref{ssec:q_chi}.  The predicted and
	observed distributions for precession \ac{snr} are in good agreement, even though the $\rho_p$ in the
	simulated signal (red lines) lies above the peak of either distribution.
	}
	\label{fig:predicted_vs_inferred_rho_p}
\end{figure}

We now apply this conjecture to a precessing signal by attempting to \emph{predict} the posterior distributions for $\rho_p$.
This allows us to investigate how much our
recovered posterior distributions may differ from the idealised case of a precession likelihood function distributed about the simulated (true) value.
In Fig.~\ref{fig:predicted_prior_rho_p} we show the results of this for the $q=4$ simulation presented in Sec.~\ref{ssec:q_chi}. This specific simulation was chosen since this case has
the largest $\rho_{p}$ and corresponds to a simulation where a non-precessing analysis is less justified.
It is therefore a good case to show how the combination of the informed prior and the additional likelihood from precession $\Lambda_{\mathrm{p}}$ correctly estimates the large $\rho_{p}$. In Fig.~\ref{fig:predicted_vs_inferred_rho_p}, we show how the predicted posterior distribution compares to the
inferred distribution over the full range of mass ratio simulations presented in Sec.~\ref{ssec:q_chi}.

In Fig.~\ref{fig:predicted_prior_rho_p} we show this predicted distribution, the informed prior, the $\chi^2$ likelihood function and the posterior distribution obtained from a full parameter estimation analysis.  By explicitly calculating the informed prior and likelihood terms separately for $\rho_p$,
we can see the effect of the prior on the $\rho_p$ posterior.
The prior strongly disfavours large
observable precession and therefore pulls the posterior towards \emph{smaller} values than the simulated value i.e. where the likelihood function peaks.

In Fig.~\ref{fig:predicted_vs_inferred_rho_p}, we show a comparison between the predicted and measured $\rho_{p}$ distributions for the set of runs with varying mass ratio presented in Sec.~\ref{ssec:q_chi}.  When we calculate the posterior, explicitly accounting for the parameter space weighting encoded in the informed prior on $\rho_{p}$, we find good agreement between the predicted and the inferred $\rho_{p}$ distributions and note that neither predicted nor inferred are centred around the \emph{true} value for the set of signals that we have simulated.  Of course, if we were to draw signals uniformly from the prior distribution, we would expect to observe the inferred distributions of $\rho_{p}$ matching with the simulated values.


\subsection{Non-precessing signal}
\label{ssec:predict_np}

\begin{figure}[t]
        \centering
        \includegraphics[scale=0.4]{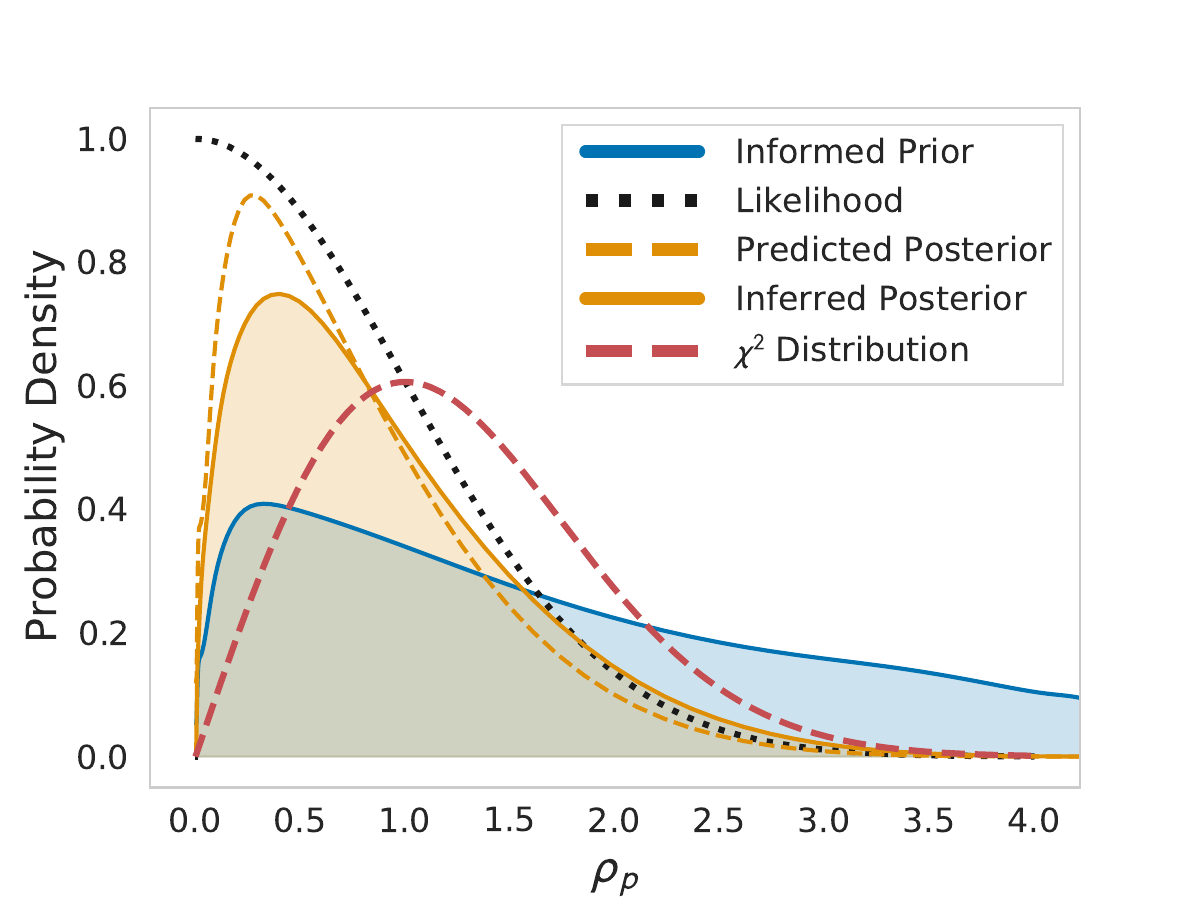}
        \caption{Distribution of $\rho_{p}$ in the absence of precession for the ``standard injection''.  The inferred $\rho_{p}$ distribution using the {\sc{IMRPhenomPv2}} approximant for recovery is shown by the solid orange line. The dashed orange line shows the predicted distribution using samples collected from an aligned-spin analysis and setting the simulated precession \ac{snr} to be 0. We also shows the $\chi^2$ distribution used previously (\cite{Fairhurst:2019_2harm}) as a red dashed line}
        \label{fig:rho_p_in_absence_of_precession}
\end{figure}

We now look at the expected posterior distribution for $\rho_p$ when there is no precession in the the signal. As explained in Sec.\ref{sec:rho_p_predict}, previously 
a $\chi^{2}$ distribution with two degrees of freedom was used to model the  $\rho_p$ distribution in the absence of any precession (see Ref.~\cite{Fairhurst:2019_2harm}). This then led to the natural heuristic that $\rho_p = 2.1$ should be the threshold for observable precession. Using Eq.~(\ref{eq:p_rho_p}) we can now use a more informative prior on $\tilde{\rho}_{p}$ and obtain a more accurate estimate of the expected posterior distribution in the absence of precession. We do this by using parameter estimation samples from an aligned-spin model and setting the simulated precession \ac{snr} to be 0, this then allows us to account for the effects of priors and different noise realisations.

In Fig.~\ref{fig:rho_p_in_absence_of_precession} we show the predicted and observed distributions for the precession \ac{snr} for a non-precessing signal.
We use a non-precessing equivalent of the ``standard'' injection as our simulated signal (i.e., we set
$\chi_{p}=0$ while ensuring all other parameters match those in Tab.~\ref{tab:standard_run}). We inject with zero noise and use the
{\texttt{IMRPhenomPv2}} model for parameter recovery.

The inferred $\rho_{p}$ distribution is peaked at lower values that the $\chi^2$ distribution as shown in Fig.~\ref{fig:rho_p_in_absence_of_precession}. However using the prediction from the likelihood (Eq.\ref{likelihood_calculation}) and the \emph{informed prior} we are able to obtain a better estimate of the posterior in the absence of precession. This estimate can be obtained without  performing parameter estimation incorporating precession, this therefore allows for a better metric for determining whether or not there is measurable precession in the system.

The distribution for the informed prior on precession \ac{snr} will depend upon the details of the signal.  In particular, it will be strongly peaked near zero for events that are likely to have small opening angle (eqivalently $\bar{b}$), i.e.,~events that are close to equal mass and have significant spin aligned with the orbital angular momentum, while high mass-ratio events and those with large anti-aligned spins will lead to greater support for large values of $\rho_p$.  Furthermore, for binaries where the orientation can be well measured, \textit{without} precession information, for example where higher modes are important, those that are close to face-on will lead to predictions of smaller $\rho_p$ while those that are edge-on will give larger values.  Given that the majority of signals observed to date are consistent with equal mass binaries, in most cases the prior on $\rho_p$ will tend to be peaked at low values.  Consequently, the simple threshold of $\rho_p \gtrsim 2.1$ as evidence for precession, remains appropriate and is likely more stringent than suggested by the simple likelihood calculation.


\section{Discussion}
\label{sec:discussion}
In most candidate astrophysical binary distributions, precession is likely to be first measured in a comparable-mass binary~\citep{Fairhurst:2019srr}.
We have considered a fiducial example of such a possible signal (mass-ratio $q=2$, \ac{snr} $\rho = 20$, and in-plane spin $\chi_p = 0.4$,
such that the precession contribution to the total \ac{snr} is $\rho_p = 5$), and performed an extensive parameter-estimation study that
has systematically explored the impact on parameter measurements of changes in each of the key source parameters:
the SNR, the in-plane spin magnitude, binary inclination, the binary mass ratio and aligned-spin contribution, the binary's total
mass, the polarisation, and sky location. These examples illustrate well-known features of precession signals~\citep{vecchio2004lisa, Lang:1900bz, Apostolatos:1994mx, Brown:2012gs, vitale2014measuring, abbott2017effects, Fairhurst:2019srr, berti2005estimating}, and
quantify their effect on both the measurement of precession, and their impact on the measurement accuracy and precision of
other parameters.

We have also verified that $\rho_p$ provides a suitable and intuitive metric for determining whether or not we have measured precession,
and shown that there is an approximate mapping between $\rho_p$ and the use of the Bayes factor to assess the evidence of precession.
We suggest that given these results, future large scale studies of precession can be made considerably computationally cheaper by
computing $\rho_p$, rather than a full Bayesian analysis.

We note that as $\rho_{p}$ captures precession by identifying additional power beyond a simple non-precessing waveform model, it could therefore be effected by phenomena such as eccentricity and higher order multipoles. As BFs simply compare the evidence for two models, one precessing and one non-precessing, using BFs as the sole metric would also be biased by properties like eccentricity and higher order multipoles. 

However, a similar approach to the 2-harmonic decomposition for precessing signals has recently been applied to
\acp{gw} including the effects of higher harmonics~\citep{mills2020measuring}.
In future work, we will combine these approaches and explore the measurability of precession
in systems with significant evidence for higher harmonics, and the impact of the combination of
higher modes and precession upon parameter accuracy. It may also be possible to account for eccentricity
through a similar decomposition.

As highlighted in section \ref{sec:rho_p_predict} these decompositions provide powerful insights into how the addition 
of physical phenomena introduce information into the analysis. Here we show that the likelihood can be simply 
factored into precessing and non-precessing contributions. This then allows us quantify the extra information that can be gained from a precessing analysis 
and even predict the recovered $\rho_p$ distribution with or without these effects taken into consideration in the analysis. 

The current study does not include higher harmonics, and uses a signal model (\texttt{IMRPhenomPv2}) that neglects two-spin precession effects,
mode asymmetries that lead to out-of-plane recoil~\cite{Kalaghatgi:2020gsq}, and detailed modelling of precession effects through merger
and ringdown. Although these effects are typically small, so is the imprint of precession on the signal, and it would be interesting in future to
investigate the impact of these additional features on our results. We also emphasize that, although we consider it to be extremely useful to
provide quantitative examples of the effects of each of the binary parameters, these will necessarily depend on the location in
parameter space of our fiducial example. However, having chosen a configuration from amongst what we expect to be the most likely
signals, we hope that these examples will act as a useful guide in interpreting precession measurements when they
arise in future gravitational-wave observations.

\section{Acknowledgements}
We are grateful to both Chris Pedersen and Alistair Muir for their initial work in this area and their useful insights. We also thank Katerina Chatziioannou and Vivien Raymond for useful discussions.
This work was supported by Science and Technology Facilities Council (STFC) grant ST/N005430/1,
and European Research Council Consolidator Grant 647839, and we are grateful for computational
resources provided by Cardiff University and supported by STFC grant ST/N000064/1.

This research has made use of data, software and/or web tools obtained from the Gravitational Wave Open Science Center (\href{https://www.gw-openscience.org}{https://www.gw-openscience.org}), a service of LIGO Laboratory, the LIGO Scientific Collaboration and the Virgo Collaboration. LIGO is funded by the U.S. National Science Foundation. Virgo is funded by the French Centre National de Recherche Scientifique (CNRS), the Italian Istituto Nazionale della Fisica Nucleare (INFN) and the Dutch Nikhef, with contributions by Polish and Hungarian institutes.

Plots were prepared with Matplotlib \citep{2007CSE.....9...90H}. The sky map
plot also used Astropy (\href{http://www.astropy.org}{http://www.astropy.org}) a community-developed core
Python package for Astronomy \citep{astropy:2013, astropy:2018} and
\texttt{ligo.skymap} (\href{https://lscsoft.docs.ligo.org/ligo.skymap}{https://lscsoft.docs.ligo.org/ligo.skymap}).
The corner plots used Corner (\href{https://corner.readthedocs.io/en/latest}{https://corner.readthedocs.io})~\citep{corner}

\bibliographystyle{unsrt_LIGO}
\bibliography{main}

\begin{thebibliography}{10}

\bibitem{Fairhurst:2019_2harm}
Stephen Fairhurst, Rhys Green, Charlie Hoy, Mark Hannam, and Alistair Muir.
\newblock {Two-harmonic approximation for gravitational waveforms from
  precessing binaries}.
\newblock {\em Phys. Rev. D}, 102:024055, 2020.

\bibitem{Fairhurst:2019srr}
Stephen Fairhurst, Rhys Green, Mark Hannam, and Charlie Hoy.
\newblock {When will we observe binary black holes precessing?}
\newblock {\em Phys. Rev. D}, 102(4):041302, 2020.

\bibitem{Abbott:2016blz}
B.~P. Abbott {\em  et~al.}
\newblock {Observation of Gravitational Waves from a Binary Black Hole Merger}.
\newblock {\em Phys. Rev. Lett.}, 116(6):061102, 2016.

\bibitem{LIGOScientific:2018mvr}
B.~P. Abbott {\em  et~al.}
\newblock {GWTC-1: A Gravitational-Wave Transient Catalog of Compact Binary
  Mergers Observed by LIGO and Virgo during the First and Second Observing
  Runs}.
\newblock {\em Phys. Rev.}, X9(3):031040, 2019.

\bibitem{Abbott:2020uma}
B.P. Abbott {\em  et~al.}
\newblock {GW190425: Observation of a Compact Binary Coalescence with Total
  Mass $\sim 3.4 M_{\odot}$}.
\newblock {\em Astrophys. J. Lett.}, 892(1):L3, 2020.

\bibitem{LIGOScientific:2020stg}
R.~Abbott {\em  et~al.}
\newblock {GW190412: Observation of a Binary-Black-Hole Coalescence with
  Asymmetric Masses}.
\newblock 4 2020.

\bibitem{Abbott:2020khf}
R.~Abbott {\em  et~al.}
\newblock {GW190814: Gravitational Waves from the Coalescence of a 23 Solar
  Mass Black Hole with a 2.6 Solar Mass Compact Object}.
\newblock {\em Astrophys. J.}, 896(2):L44, 2020.

\bibitem{Abbott:2020mjq}
R.~Abbott {\em  et~al.}
\newblock {Properties and astrophysical implications of the 150 Msun binary
  black hole merger GW190521}.
\newblock {\em Astrophys. J. Lett.}, 900:L13, 2020.

\bibitem{Abbott:2020tfl}
R.~Abbott {\em  et~al.}
\newblock {GW190521: A Binary Black Hole Merger with a Total Mass of $150 ~
  M_{\odot}$}.
\newblock {\em Phys. Rev. Lett.}, 125(10):101102, 2020.

\bibitem{nitz20202}
Alexander~H Nitz, {\em  et~al.}
\newblock 2-ogc: Open gravitational-wave catalog of binary mergers from
  analysis of public advanced ligo and virgo data.
\newblock {\em The Astrophysical Journal}, 891(2):123, 2020.

\bibitem{venumadhav2020new}
Tejaswi Venumadhav, Barak Zackay, Javier Roulet, Liang Dai, and Matias
  Zaldarriaga.
\newblock New binary black hole mergers in the second observing run of advanced
  ligo and advanced virgo.
\newblock {\em Physical Review D}, 101(8):083030, 2020.

\bibitem{zackay2019highly}
Barak Zackay, Tejaswi Venumadhav, Liang Dai, Javier Roulet, and Matias
  Zaldarriaga.
\newblock Highly spinning and aligned binary black hole merger in the advanced
  ligo first observing run.
\newblock {\em Physical Review D}, 100(2):023007, 2019.

\bibitem{zackay2019detecting}
Barak Zackay, Liang Dai, Tejaswi Venumadhav, Javier Roulet, and Matias
  Zaldarriaga.
\newblock Detecting gravitational waves with disparate detector responses: two
  new binary black hole mergers.
\newblock {\em arXiv preprint arXiv:1910.09528}, 2019.

\bibitem{SchutzDeterminingHubbleconstant1986}
Bernard~F. Schutz.
\newblock Determining the {{Hubble}} constant from gravitational wave
  observations.
\newblock {\em Nature}, 323(6086):310--311, September 1986.

\bibitem{Soares-Santos:2019irc}
M.~Soares-Santos {\em  et~al.}
\newblock {First measurement of the Hubble constant from a dark standard siren
  using the Dark Energy Survey galaxies and the LIGO/Virgo binary-black-hole
  merger GW170814}.
\newblock {\em Submitted to: Astrophys. J.}, 2019.

\bibitem{AbbottGravitationalWavesGammaRays2017}
B.~P. Abbott, {\em  et~al.}
\newblock Gravitational {{Waves}} and {{Gamma}}-{{Rays}} from a {{Binary
  Neutron Star Merger}}: {{GW170817}} and {{GRB 170817A}}.
\newblock {\em The Astrophysical Journal}, 848(2):L13, October 2017.

\bibitem{abbott2018gw170817}
Benjamin~P Abbott, {\em  et~al.}
\newblock Gw170817: Measurements of neutron star radii and equation of state.
\newblock {\em Physical review letters}, 121(16):161101, 2018.

\bibitem{abbott2018gw170817stochastic}
Benjamin~P Abbott, {\em  et~al.}
\newblock Gw170817: implications for the stochastic gravitational-wave
  background from compact binary coalescences.
\newblock {\em Physical review letters}, 120(9):091101, 2018.

\bibitem{ligo2017gravitational}
LIGO~Scientific Collaboration, {\em  et~al.}
\newblock A gravitational-wave standard siren measurement of the hubble
  constant.
\newblock {\em Nature}, 551(7678):85--88, 2017.

\bibitem{abbott2019tests}
BP~Abbott, {\em  et~al.}
\newblock Tests of general relativity with the binary black hole signals from
  the ligo-virgo catalog gwtc-1.
\newblock {\em Physical Review D}, 100(10):104036, 2019.

\bibitem{LIGOScientific:2018jsj}
B.~P. Abbott {\em  et~al.}
\newblock {Binary Black Hole Population Properties Inferred from the First and
  Second Observing Runs of Advanced LIGO and Advanced Virgo}.
\newblock {\em Astrophys. J.}, 882(2):L24, 2019.

\bibitem{Aasi:2013wya}
B.P. Abbott {\em  et~al.}
\newblock {Prospects for Observing and Localizing Gravitational-Wave Transients
  with Advanced LIGO, Advanced Virgo and KAGRA}.
\newblock {\em Living Rev. Rel.}, 21(1):3, 2018.

\bibitem{TheLIGOScientific:2014jea}
J.~Aasi {\em  et~al.}
\newblock {Advanced LIGO}.
\newblock {\em Class. Quant. Grav.}, 32:074001, 2015.

\bibitem{acernese2014advanced}
F~Acernese, {\em  et~al.}
\newblock Advanced virgo: a second-generation interferometric gravitational
  wave detector.
\newblock {\em Classical and Quantum Gravity}, 32(2):024001, 2014.

\bibitem{aso2013interferometer}
Yoichi Aso, {\em  et~al.}
\newblock Interferometer design of the kagra gravitational wave detector.
\newblock {\em Physical Review D}, 88(4):043007, 2013.

\bibitem{Sathyaprakash:2019nnu}
B.~S. Sathyaprakash {\em  et~al.}
\newblock {Cosmology and the Early Universe}.
\newblock 2019.

\bibitem{Bianchi:2018ula}
Eugenio Bianchi, Anuradha Gupta, Hal~M. Haggard, and B.~S. Sathyaprakash.
\newblock {Quantum gravity and black hole spin in gravitational wave
  observations: a test of the Bekenstein-Hawking entropy}.
\newblock 2018.

\bibitem{Ford:2019nic}
K.~E.~Saavik Ford, {\em  et~al.}
\newblock {Multi-Messenger Astrophysics Opportunities with Stellar-Mass Binary
  Black Hole Mergers}.
\newblock 2019.

\bibitem{Cutler:1994ys}
Curt Cutler and Eanna~E. Flanagan.
\newblock {Gravitational waves from merging compact binaries: How accurately
  can one extract the binary's parameters from the inspiral wave form?}
\newblock {\em Phys. Rev.}, D49:2658--2697, 1994.

\bibitem{LIGOScientificCollaborationandVirgoCollaborationPropertiesBinaryBlack2016}
{LIGO Scientific Collaboration and Virgo Collaboration}, {\em  et~al.}
\newblock Properties of the {{Binary Black Hole Merger GW150914}}.
\newblock {\em Phys. Rev. Lett.}, 116(24):241102, June 2016.

\bibitem{Usman:2018imj}
Samantha~A. Usman, Joseph~C. Mills, and Stephen Fairhurst.
\newblock {Constraining the Inclinations of Binary Mergers from
  Gravitational-wave Observations}.
\newblock {\em Astrophys. J.}, 877(2):82, 2019.

\bibitem{Poisson:1995ef}
Eric Poisson and Clifford~M. Will.
\newblock {Gravitational waves from inspiraling compact binaries: Parameter
  estimation using second postNewtonian wave forms}.
\newblock {\em Phys. Rev.}, D52:848--855, 1995.

\bibitem{Baird:2012cu}
Emily Baird, Stephen Fairhurst, Mark Hannam, and Patricia Murphy.
\newblock {Degeneracy between mass and spin in black-hole-binary waveforms}.
\newblock {\em Phys. Rev.}, D87(2):024035, 2013.

\bibitem{farr2016parameter}
Ben Farr, {\em  et~al.}
\newblock Parameter estimation on gravitational waves from neutron-star
  binaries with spinning components.
\newblock {\em The Astrophysical Journal}, 825(2):116, 2016.

\bibitem{ng2018gravitational}
Ken~KY Ng, {\em  et~al.}
\newblock Gravitational-wave astrophysics with effective-spin measurements:
  Asymmetries and selection biases.
\newblock {\em Physical Review D}, 98(8):083007, 2018.

\bibitem{Graff:2015bba}
Philip~B. Graff, Alessandra Buonanno, and B.~S. Sathyaprakash.
\newblock {Missing Link: Bayesian detection and measurement of
  intermediate-mass black-hole binaries}.
\newblock {\em Phys. Rev. D}, 92(2):022002, 2015.

\bibitem{Haster:2015cnn}
Carl-Johan Haster, {\em  et~al.}
\newblock {Inference on gravitational waves from coalescences of stellar-mass
  compact objects and intermediate-mass black holes}.
\newblock {\em Mon. Not. Roy. Astron. Soc.}, 457(4):4499--4506, 2016.

\bibitem{Vitale:2016avz}
Salvatore Vitale, {\em  et~al.}
\newblock {Parameter estimation for heavy binary-black holes with networks of
  second-generation gravitational-wave detectors}.
\newblock {\em Phys. Rev. D}, 95(6):064053, 2017.

\bibitem{Yu:2017zgi}
Hang Yu {\em  et~al.}
\newblock {Prospects for detecting gravitational waves at 5 Hz with
  ground-based detectors}.
\newblock {\em Phys. Rev. Lett.}, 120(14):141102, 2018.

\bibitem{Apostolatos:1994mx}
Theocharis~A. Apostolatos, Curt Cutler, Gerald~J. Sussman, and Kip~S. Thorne.
\newblock {Spin induced orbital precession and its modulation of the
  gravitational wave forms from merging binaries}.
\newblock {\em Phys. Rev.}, D49:6274--6297, 1994.

\bibitem{Kidder:1995zr}
Lawrence~E. Kidder.
\newblock {Coalescing binary systems of compact objects to postNewtonian 5/2
  order. 5. Spin effects}.
\newblock {\em Phys. Rev.}, D52:821--847, 1995.

\bibitem{Hannam:2013oca}
Mark Hannam, {\em  et~al.}
\newblock {Simple Model of Complete Precessing Black-Hole-Binary Gravitational
  Waveforms}.
\newblock {\em Phys. Rev. Lett.}, 113(15):151101, 2014.

\bibitem{Pan:2013rra}
Yi~Pan, {\em  et~al.}
\newblock {Inspiral-merger-ringdown waveforms of spinning, precessing
  black-hole binaries in the effective-one-body formalism}.
\newblock {\em Phys. Rev. D}, 89(8):084006, 2014.

\bibitem{abbott2019open}
R~Abbott, {\em  et~al.}
\newblock Open data from the first and second observing runs of advanced ligo
  and advanced virgo.
\newblock {\em arXiv preprint arXiv:1912.11716}, 2019.

\bibitem{vecchio2004lisa}
Alberto Vecchio.
\newblock Lisa observations of rapidly spinning massive black hole binary
  systems.
\newblock {\em Physical Review D}, 70(4):042001, 2004.

\bibitem{Lang:1900bz}
Ryan~N. Lang and Scott~A. Hughes.
\newblock {Measuring coalescing massive binary black holes with gravitational
  waves: The Impact of spin-induced precession}.
\newblock {\em Phys. Rev.}, D74:122001, 2006.
\newblock [Erratum: Phys. Rev.D77,109901(2008)].

\bibitem{Chatziioannou:2018wqx}
Katerina Chatziioannou, {\em  et~al.}
\newblock {Measuring the properties of nearly extremal black holes with
  gravitational waves}.
\newblock {\em Phys. Rev. D}, 98(4):044028, 2018.

\bibitem{Brown:2012gs}
Duncan~A. Brown, Andrew Lundgren, and R.~O'Shaughnessy.
\newblock {Nonspinning searches for spinning binaries in ground-based detector
  data: Amplitude and mismatch predictions in the constant precession cone
  approximation}.
\newblock {\em Phys. Rev.}, D86:064020, 2012.

\bibitem{vitale2014measuring}
Salvatore Vitale, Ryan Lynch, John Veitch, Vivien Raymond, and Riccardo
  Sturani.
\newblock Measuring the spin of black holes in binary systems using
  gravitational waves.
\newblock {\em Physical Review Letters}, 112(25):251101, 2014.

\bibitem{abbott2017effects}
Benjamin~P Abbott, {\em  et~al.}
\newblock Effects of waveform model systematics on the interpretation of
  gw150914.
\newblock {\em Classical and Quantum Gravity}, 34(10):104002, 2017.

\bibitem{berti2005estimating}
Emanuele Berti, Alessandra Buonanno, and Clifford~M Will.
\newblock Estimating spinning binary parameters and testing alternative
  theories of gravity with lisa.
\newblock {\em Physical Review D}, 71(8):084025, 2005.

\bibitem{Pratten:2020igi}
Geraint Pratten, Patricia Schmidt, Riccardo Buscicchio, and Lucy~M. Thomas.
\newblock {On measuring precession in GW190814-like asymmetric compact
  binaries}.
\newblock 6 2020.

\bibitem{Schmidt:2014iyl}
Patricia Schmidt, Frank Ohme, and Mark Hannam.
\newblock {Towards models of gravitational waveforms from generic binaries II:
  Modelling precession effects with a single effective precession parameter}.
\newblock {\em Phys. Rev.}, D91(2):024043, 2015.

\bibitem{Gerosa:2020aiw}
Davide Gerosa, {\em  et~al.}
\newblock {A generalized precession parameter $\chi_{\rm p}$ to interpret
  gravitational-wave data}.
\newblock 11 2020.

\bibitem{Thomas:2020uqj}
Lucy~M. Thomas, Patricia Schmidt, and Geraint Pratten.
\newblock {A new effective precession spin for modelling multi-modal
  gravitational waveforms in the strong-field regime}.
\newblock 12 2020.

\bibitem{Abbott:2020niy}
R.~Abbott {\em  et~al.}
\newblock {GWTC-2: Compact Binary Coalescences Observed by LIGO and Virgo
  During the First Half of the Third Observing Run}.
\newblock 10 2020.

\bibitem{Abbott:2020gyp}
R.~Abbott {\em  et~al.}
\newblock {Population Properties of Compact Objects from the Second LIGO-Virgo
  Gravitational-Wave Transient Catalog}.
\newblock 10 2020.

\bibitem{LIGOScientificCollaborationandVirgoCollaborationGW151226ObservationGravitational2016}
{LIGO Scientific Collaboration and Virgo Collaboration}, {\em  et~al.}
\newblock {{GW151226}}: {{Observation}} of {{Gravitational Waves}} from a
  22-{{Solar}}-{{Mass Binary Black Hole Coalescence}}.
\newblock {\em Phys. Rev. Lett.}, 116(24):241103, June 2016.

\bibitem{Farr:2017uvj}
Will~M. Farr, {\em  et~al.}
\newblock {Distinguishing Spin-Aligned and Isotropic Black Hole Populations
  With Gravitational Waves}.
\newblock {\em Nature}, 548:426, 2017.

\bibitem{Tiwari:2018qch}
Vaibhav Tiwari, Stephen Fairhurst, and Mark Hannam.
\newblock {Constraining black-hole spins with gravitational wave observations}.
\newblock {\em Astrophys. J.}, 868(2):140, 2018.

\bibitem{Buonanno:2004yd}
Alessandra Buonanno, Yan-bei Chen, Yi~Pan, and Michele Vallisneri.
\newblock {A Quasi-physical family of gravity-wave templates for precessing
  binaries of spinning compact objects. 2. Application to double-spin
  precessing binaries}.
\newblock {\em Phys. Rev. D}, 70:104003, 2004.
\newblock [Erratum: Phys.Rev.D 74, 029902 (2006)].

\bibitem{Bohe:PhenomPv2}
{Boh{\'e}, Alejandro and Hannam, Mark and Husa, Sascha and Ohme, Frank and
  Puerrer, Michael and Schmidt, Patricia}.
\newblock Phenompv2 - technical notes for lal implementation.
\newblock Technical Report {LIGO}-T1500602, {LIGO} Project, 2016.

\bibitem{abbott2017gw170814}
Benjamin~P Abbott, {\em  et~al.}
\newblock Gw170814: a three-detector observation of gravitational waves from a
  binary black hole coalescence.
\newblock {\em Physical review letters}, 119(14):141101, 2017.

\bibitem{buikema2020sensitivity}
A~Buikema, {\em  et~al.}
\newblock Sensitivity and performance of the advanced ligo detectors in the
  third observing run.
\newblock {\em Physical Review D}, 102(6):062003, 2020.

\bibitem{abbott2018prospects}
Benjamin~P Abbott, {\em  et~al.}
\newblock Prospects for observing and localizing gravitational-wave transients
  with advanced ligo, advanced virgo and kagra.
\newblock {\em Living Reviews in Relativity}, 21(1):3, 2018.

\bibitem{finn1992detection}
Lee~S Finn.
\newblock Detection, measurement, and gravitational radiation.
\newblock {\em Physical Review D}, 46(12):5236, 1992.

\bibitem{Veitch:2014wba}
J.~Veitch {\em  et~al.}
\newblock {Parameter estimation for compact binaries with ground-based
  gravitational-wave observations using the LALInference software library}.
\newblock {\em Phys. Rev.}, D91(4):042003, 2015.

\bibitem{LALSuite}
{LIGO Scientific Collaboration}.
\newblock {LIGO Algorithm Library}, 2018.

\bibitem{Khan:2015jqa}
Sebastian Khan, {\em  et~al.}
\newblock {Frequency-domain gravitational waves from nonprecessing black-hole
  binaries. II. A phenomenological model for the advanced detector era}.
\newblock {\em Phys. Rev.}, D93(4):044007, 2016.

\bibitem{Husa:2015iqa}
Sascha Husa, {\em  et~al.}
\newblock {Frequency-domain gravitational waves from nonprecessing black-hole
  binaries. I. New numerical waveforms and anatomy of the signal}.
\newblock {\em Phys. Rev.}, D93(4):044006, 2016.

\bibitem{Hoy:2020vys}
Charlie Hoy and Vivien Raymond.
\newblock {PESummary: the code agnostic Parameter Estimation Summary page
  builder}.
\newblock 6 2020.

\bibitem{kalaghatgi2019parameter}
Chinmay Kalaghatgi, Mark Hannam, and Vivien Raymond.
\newblock Parameter estimation with a spinning multi-mode waveform model:
  Imrphenomhm.
\newblock {\em arXiv preprint arXiv:1909.10010}, 2019.

\bibitem{vitale2018measuring}
Salvatore Vitale and Hsin-Yu Chen.
\newblock Measuring the hubble constant with neutron star black hole mergers.
\newblock {\em Physical review letters}, 121(2):021303, 2018.

\bibitem{Fairhurst:2009tc}
Stephen Fairhurst.
\newblock {Triangulation of gravitational wave sources with a network of
  detectors}.
\newblock {\em New J. Phys.}, 11:123006, 2009.
\newblock [Erratum: New J. Phys.13,069602(2011)].

\bibitem{Singer:2015ema}
Leo~P. Singer and Larry~R. Price.
\newblock {Rapid Bayesian position reconstruction for gravitational-wave
  transients}.
\newblock {\em Phys. Rev.}, D93(2):024013, 2016.

\bibitem{Vallisneri:2007ev}
Michele Vallisneri.
\newblock {Use and abuse of the Fisher information matrix in the assessment of
  gravitational-wave parameter-estimation prospects}.
\newblock {\em Phys. Rev. D}, 77:042001, 2008.

\bibitem{OShaughnessy:2014shr}
R.~O'Shaughnessy, {\em  et~al.}
\newblock {Parameter estimation of gravitational waves from precessing black
  hole-neutron star inspirals with higher harmonics}.
\newblock {\em Phys. Rev.}, D89(10):102005, 2014.

\bibitem{Campanelli:2006uy}
Manuela Campanelli, C.O. Lousto, and Y.~Zlochower.
\newblock {Spinning-black-hole binaries: The orbital hang up}.
\newblock {\em Phys. Rev. D}, 74:041501, 2006.

\bibitem{mills2020measuring}
Cameron Mills and Stephen Fairhurst.
\newblock Measuring gravitational-wave higher-order modes.
\newblock {\em arXiv preprint arXiv:2007.04313}, 2020.

\bibitem{jaynes2003probability}
Edwin~T Jaynes.
\newblock {\em Probability theory: The logic of science}.
\newblock Cambridge university press, 2003.

\bibitem{Veitch:2009hd}
J.~Veitch and A.~Vecchio.
\newblock {Bayesian coherent analysis of in-spiral gravitational wave signals
  with a detector network}.
\newblock {\em Phys. Rev.}, D81:062003, 2010.

\bibitem{cornish2011gravitational}
Neil Cornish, Laura Sampson, Nicolas Yunes, and Frans Pretorius.
\newblock Gravitational wave tests of general relativity with the parameterized
  post-einsteinian framework.
\newblock {\em Physical Review D}, 84(6):062003, 2011.

\bibitem{Kalaghatgi:2020gsq}
Chinmay Kalaghatgi and Mark Hannam.
\newblock {Investigating the effect of in-plane spin directions for Precessing
  BBH systems}.
\newblock 8 2020.

\bibitem{2007CSE.....9...90H}
J.~D. {Hunter}.
\newblock {Matplotlib: A 2D Graphics Environment}.
\newblock {\em CSE}, 9:90--95, May 2007.

\bibitem{astropy:2013}
{Astropy Collaboration}, {\em  et~al.}
\newblock {Astropy: A community Python package for astronomy}.
\newblock {\em Astronomy and Astrophysics}, 558:A33, October 2013.

\bibitem{astropy:2018}
A.~M. {Price-Whelan}, {\em  et~al.}
\newblock {The Astropy Project: Building an Open-science Project and Status of
  the v2.0 Core Package}.
\newblock {\em The Astronomical Journal}, 156:123, September 2018.

\bibitem{corner}
Daniel Foreman-Mackey.
\newblock corner.py: Scatterplot matrices in python.
\newblock {\em The Journal of Open Source Software}, 1(2):24, jun 2016.

\end{thebibliography}

\end{document}